\documentclass[a4paper,12pt]{article}
\pdfoutput=1

\usepackage{mathrsfs,graphicx,rotating,amsmath,amsfonts,mathtools,booktabs,amssymb}
\usepackage{hyperref}\usepackage{slashed}
\usepackage[nosort]{cite}
\usepackage[table,xcdraw]{xcolor}
\usepackage{youngtab}
\usepackage{graphicx}
\usepackage{multirow,multicol}
\hypersetup{colorlinks,bookmarksopen,bookmarksnumbered,
linkcolor=blus,pdfstartview=FitH,urlcolor=rossos,citecolor=verde}
\allowdisplaybreaks

\newcommand{\mio}[1]{}

 \newcommand{\med}[1]{\langle #1\rangle}

 \newcommand{\fig}[1]{~\ref{fig:#1}}
\newcommand{\sfrac}[2]{#1/#2} 
 \newcommand{\One}{1\!\!\hbox{I}}
\newcommand{\MDM}{m_\Q}

\newcommand{\Q}{{\cal Q}}
\newcommand{\B}{{\cal B}}

\allowdisplaybreaks
\usepackage{multicol}
\usepackage{color}
\definecolor{rosso}{cmyk}{0,1,1,0.4}
\definecolor{rossos}{cmyk}{0,1,1,0.55}
\definecolor{rossoc}{cmyk}{0,1,1,0.2}
\definecolor{blu}{cmyk}{1,1,0,0.3}
\definecolor{blus}{cmyk}{1,1,0,0.6}
\definecolor{bluc}{cmyk}{1,1,0,0.1}
\definecolor{verde}{cmyk}{0.92,0,0.59,0.25}
\definecolor{verdec}{cmyk}{0.92,0,0.59,0.15}
\definecolor{verdes}{cmyk}{0.92,0,0.59,0.4}

\newcommand{\eq}[1]{~{\rm (\ref{eq:#1})}}

\newcommand{\KeV}{\,{\rm KeV}}
\newcommand{\MeV}{\,{\rm MeV}}
\newcommand{\GeV}{\,{\rm GeV}}
\newcommand{\TeV}{\,{\rm TeV}}
\newcommand{\cm}{\,{\rm cm}}
\newcommand{\Tr}{\,{\rm Tr}}

\def\circa#1{\,\raise.3ex\hbox{$#1$\kern-.75em\lower1ex\hbox{$\sim$}}\,}

\newcommand{\beq}{\begin{equation}}
\newcommand{\eeq}{\end{equation}}
\newcommand{\mb}[1]{\mbox{\boldmath $#1$}}

\newcommand{\bea}{\begin{eqnarray}}
\newcommand{\eea}{\end{eqnarray}}
\newcommand{\be}{\begin{equation}}
\newcommand{\ee}{\end{equation}}
\font\tenrsfs=rsfs10 at 12pt
\font\sevenrsfs=rsfs7
\font\fiversfs=rsfs5
\newfam\rsfsfam
\textfont\rsfsfam=\tenrsfs
\scriptfont\rsfsfam=\sevenrsfs
\scriptscriptfont\rsfsfam=\fiversfs

\newcommand{\La}{\mathscr{L}}

\newsavebox\MBox

\newcommand{\SOD}{\SO(\ndc)}
\newcommand{\LDC}{\Lambda_{\rm DC}}
\newcommand{\tauDG}{\tau_{\scriptscriptstyle{\rm DG}}}
\newcommand{\mDG}{M_{\scriptscriptstyle{\rm DG}}}
\newcommand{\GammaDG}{\Gamma_{\scriptscriptstyle{\rm DG}}}

\newcommand{\mQ}{m_{\mathcal{Q}}}

\newcommand{\adc}{\alpha_{\scriptscriptstyle{\rm DC}}}

\newcommand{\eV}{\,{\rm eV}}
\newcommand{\SU}{\,{\rm SU}}
\newcommand{\SO}{\,{\rm SO}}
\newcommand{\U}{\,{\rm U}}

\def\circa#1{\,\raise.3ex\hbox{$#1$\kern-.75em\lower1ex\hbox{$\sim$}}\,}
\makeatletter

\font\ital=cmu10 

\def\hhref#1{\href{http://arxiv.org/abs/#1}{arXiv:#1}}
\usepackage{xstring} 
\newcommand{\hhrefq}[1]{\IfSubStr{#1}{:}{\href{http://inspirehep.net/search?ln=en&ln=en&p=#1&of=hb&action_search=Search&sf=&so=d&rm=&rg=25&sc=0}{InSpires:#1}}{\hhref{#1}}}

\def\art{\@ifnextchar[{\eart}{\oart}}
\def\eart[#1]#2#3#4#5#6{{\rm #2}, {\em #3 \bf #4} {\rm (#6) #5} ({\em #1})}
\def\article{\@ifnextchar[{\earticle}{\oarticle}}
\def\oarticle#1#2#3#4#5#6{{\rm #1}, {\ital ``#6''}, {\rm #2 #3 (#5) #4}}
\def\earticle[#1]#2#3#4#5#6#7{{\rm #2}, {\ital ``#7''}, {\rm #3 #4 (#6) #5}  [\hhrefq{#1}]}
\def\hepart[#1]#2{{\rm #2, \sl#1}}
\def\heparticle[#1]#2#3{#2, {\ital ``#3''} [\hhrefq{#1}]}
\newcommand{\doi}[1]{\href{http://dx.doi.org/#1}{[link]}}

\renewenvironment{thebibliography}[1]
     {\begin{multicols}{2}[\section*{\refname}]%
      \@mkboth{\MakeUppercase\refname}{\MakeUppercase\refname}%
      \list{\@biblabel{\@arabic\c@enumiv}}%
           {\settowidth\labelwidth{\@biblabel{#1}}%
            \leftmargin\labelwidth
            \advance\leftmargin\labelsep
            \@openbib@code
            \usecounter{enumiv}%
            \let\p@enumiv\@empty
            \renewcommand\theenumiv{\@arabic\c@enumiv}}%
      \sloppy
      \clubpenalty4000
      \@clubpenalty \clubpenalty
      \widowpenalty4000%
      \sfcode`\.\@m}
     {\def\@noitemerr
       {\@latex@warning{Empty `thebibliography' environment}}%
      \endlist\end{multicols}}

\newcounter{alphaequation}[equation]
\def\thealphaequation{\theequation\hbox to
0.6em{\hfil\alph{alphaequation}\hfil}}
\def\eqnsystem#1{
\def\@eqnnum{{\rm (\thealphaequation)}}
\def\@@eqncr{\let\@tempa\relax \ifcase\@eqcnt \def\@tempa{& & &} \or
  \def\@tempa{& &}\or \def\@tempa{&}\fi\@tempa
  \if@eqnsw\@eqnnum\refstepcounter{alphaequation}\fi
\global\@eqnswtrue\global\@eqcnt=0\cr}
\refstepcounter{equation} \let\@currentlabel\theequation \def\@tempb{#1}
\ifx\@tempb\empty\else\label{#1}\fi
\refstepcounter{alphaequation}
\let\@currentlabel\thealphaequation
\global\@eqnswtrue\global\@eqcnt=0 \tabskip\@centering\let\\=\@eqncr
$$\halign to \displaywidth\bgroup \@eqnsel\hskip\@centering
$\displaystyle\tabskip\z@{##}$&\global\@eqcnt\@ne
\hskip2\arraycolsep\hfil${##}$\hfil& \global\@eqcnt\tw@\hskip2\arraycolsep
$\displaystyle\tabskip\z@{##}$\hfil
\tabskip\@centering&\llap{##}\tabskip\z@\cr}
\def\endeqnsystem{\@@eqncr\egroup$$\global\@ignoretrue} \makeatother

\newcommand{\ndc}{N_{\rm DC}}

\definecolor{Gray}{gray}{0.95}

\begin{document}

{CERN-TH-2017-151\hfill IFUP-TH/2017}

\vspace{2cm}

\begin{center}
{\Large\LARGE \bf \color{rossos}
Dark Matter as a weakly coupled Dark Baryon
}\\[1cm]
{\bf Andrea Mitridate$^{a}$, Michele Redi$^{b}$, Juri Smirnov$^{b}$, Alessandro Strumia$^{c,d}$}  
\\[7mm]

{\it $^a$Scuola Normale Superiore and INFN, Pisa, Italy}\\[1mm]
{\it $^b$ INFN and Department of Physics and Astronomy\\ University of Florence,
Via G. Sansone 1, 50019 Sesto Fiorentino, Italy}\\[1mm]
{\it $^c$ Dipartimento di Fisica dell'Universit{\`a} di Pisa and INFN, Italy}\\[1mm]
{\it $^d$ CERN, Theory Division, Geneva, Switzerland}\\[1mm]

\vspace{2cm}

{\large\bf\color{blus} Abstract}
\begin{quote}
Dark Matter might be an accidentally stable baryon of a new confining gauge interaction.
We extend previous studies exploring the possibility that the DM is made of dark quarks
heavier than the dark confinement scale. The resulting phenomenology contains new unusual elements:
a two-stage DM cosmology (freeze-out followed by dark condensation),
a large DM annihilation cross section through recombination of dark quarks
(allowing to fit the positron excess). Light dark glue-balls are relatively long lived and give extra cosmological effects;
DM itself can remain radioactive.
\end{quote}

\thispagestyle{empty}
\bigskip\newpage

\end{center}
\begin{quote}
{\large\noindent\color{blus} 
}

\end{quote}
\vspace{-1.5cm}

\tableofcontents

\setcounter{footnote}{0}

\section{Introduction}
Dark Matter (DM) could be a new massive particle, neutral and stable on cosmological time scales.
In the absence of experimental indications, so many models of particle DM have been proposed
that discussing one more possibility risks of being superfluous.

In this paper we explore the possibility that DM is a dark-baryon, made of $\ndc$ copies of a dark-quark $\Q$
with mass $m_\Q$,
much larger than the scale $\LDC$ where a new dark-color gauge interaction becomes strong.
We believe that this possibility deserves to be studied because of the following elements of interest.

This simple and predictive 
scenario explains DM stability in the same way in which the Standard Model (SM) explains proton stability.
DM is stable because the renormalizable theory has an accidental symmetry, dark-baryon number.
No ad-hoc symmetry (such as $R$-parity or Z$_2$) needs to be imposed by hand.

The systematic study of such scenarios was initiated in~\cite{Antipin:2015xia}, 
where dark quarks were assumed to be lighter than  the confinement scale $\LDC$ of the gauge theory, see also \cite{0604261,0909.2034 ,1410.1817,1402.6656,1506.06929,1607.07865,1604.04627,1706.02722}.  
 In this work we  explore the opposite regime  with heavy fermions, {see also \cite{1606.00159}}. 
 This leads to increased predictivity: in the presence of multiple dark-quarks,
 only the lightest one is typically relevant for DM physics, that is thereby determined in terms of
 two free parameters, $m_\Q$ and $\LDC$.
 
\begin{figure}[!t]
\begin{center}
\includegraphics[width=.5\textwidth]{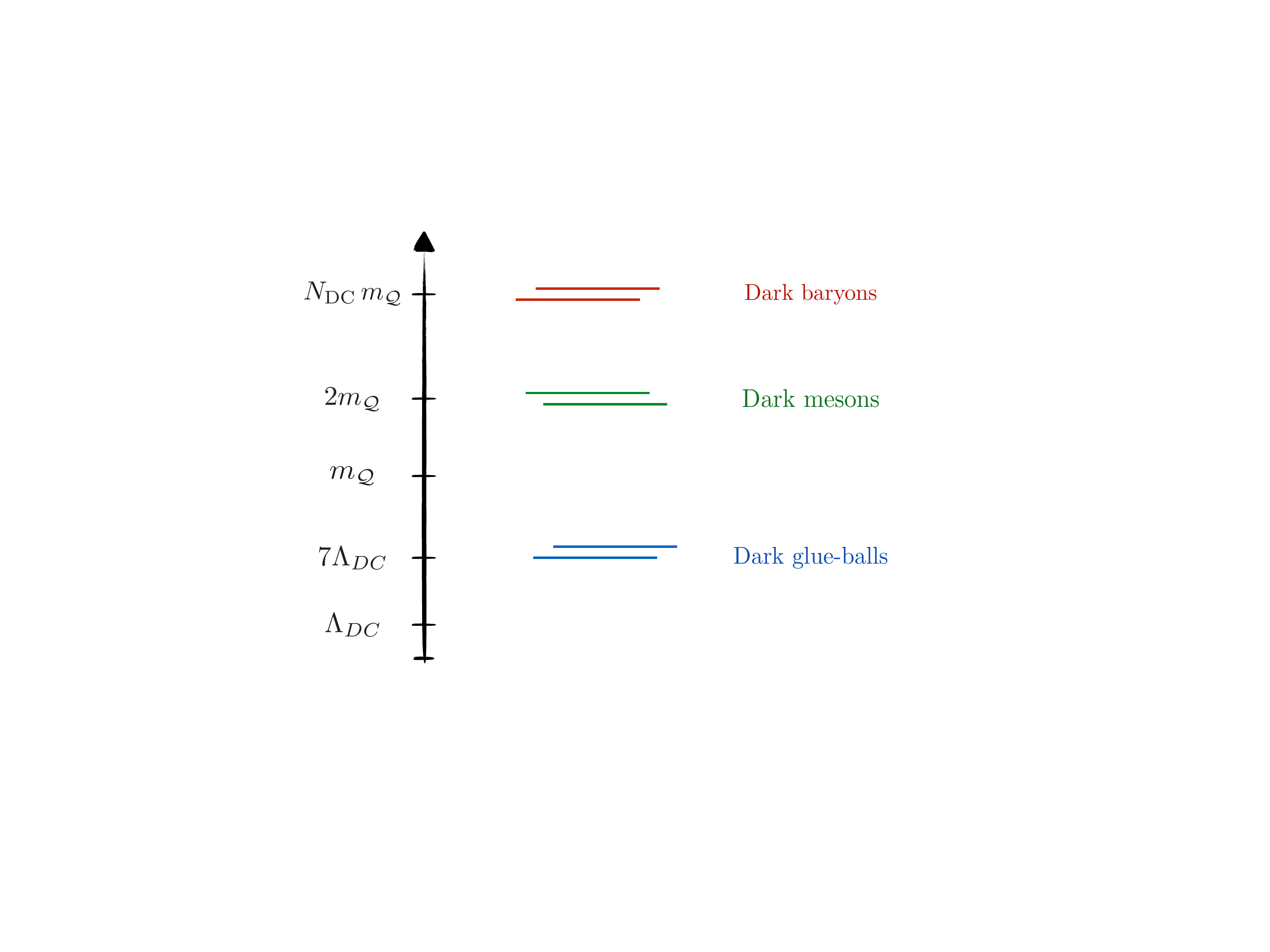}\\
\caption{\em Typical spectrum of the theory. We assume that the lightest dark quark is heavier than the dark confinement scale, 
$\LDC$.  DM is dark baryon made of $N_{\rm DC}$  dark quarks. 
The lightest dark states are unstable dark glue-balls.}
\label{fig:spectrum}
\end{center}
\end{figure}

Furthermore, it leads to novel characteristic signatures.
 \begin{enumerate}
\item  The cosmological  history is not standard, and the relic
DM abundance is determined in two stages:
the dark-quark relic abundance freezes out at  $T \sim m_\Q/25$ in the usual way,
through weakly coupled annihilations with cross section
$\sigma_{\Q\bar\Q} v_{\rm rel}\sim \pi \adc^2/m_\Q^2$.
This is followed at $T \sim \LDC$ by a first-order dark phase transition~\cite{firstorder}, where a fraction of the dark quarks $\Q$ and $\bar \Q$
binds into mesons, that decay, and the remaining fraction forms stable dark-matter baryons $\B$ and $\bar \B$.

\item The $\B \bar \B$ annihilation cross section relevant for indirect DM detection
is a few orders of magnitude larger than the usual $\Q\bar\Q$ annihilation cross section, 
being enhanced by dark-atomic $1/\adc$ effects.

\item Fig.~\ref{fig:spectrum} illustrates the spectrum of the theory:
the dark sector contains unstable dark-glue-balls with mass $\mDG \sim \LDC$
which can be much lighter than DM with mass $\sim m_\Q$,
and thereby potentially accessible  to low-energy searches, such as high-luminosity fixed-target experiments. 
If $\mDG$ is larger than the binding energy,
some dark quarks could have formed long-lived excited dark baryons, that de-excite emitting $\beta$ or $\gamma$ radio-activity.


 \end{enumerate}
The paper is organized as follows. In section~\ref{2} we outline the scenario and the main options:
$\SU(\ndc)$ and $\SO(\ndc)$ gauge theories, with dark quarks neutral or charged under the SM gauge group.
In section~\ref{bound} we study the bound states: lighter unstable dark glue-balls, dark mesons, stable dark baryons;
we compute their binding energies by means of a variational method.
In section~\ref{relic} we study how baryon DM can form throughout  the cosmological history.
In section~\ref{Signatures} we study signatures in 
cosmology, direct detection, indirect detection (enhanced by recombination), colliders, high-intensity experiments at lower energy,
radioactive DM.
Detailed computations in the main specific models are presented in section~\ref{Models}.
In section~\ref{Conclusions} we conclude summarising the main novel results. 

\section{The scenario}\label{2}
We consider DM made of `dark quarks', new fermions possibly charged under the SM gauge group and
charged under a new confining gauge interaction $G_{\rm DC} =\SU(\ndc)$ or $\SO(\ndc)$.
We will dub the new interaction Dark Color (DC).
The dark-quarks are assumed to lie in the fundamental representation of the DC group and to form a vectorial representation $R$ (in general reducible) of the SM 
\begin{equation}
\Q\equiv  (\ndc,R) \oplus  (\bar N_{\rm DC},\overline{R})
\end{equation}
where $\ndc$ and $\bar N_{\rm DC}$ indicate respectively the fundamental and anti-fundamental representation of the dark-color group, 
and $R$ is a representation of the SM groups.
These theories are described by the  renormalizable Lagrangian
\be
\La=\La_{\rm SM}-\frac{1}{4g_{\rm DC}^2}\mathcal{G}^A_{\mu\nu}\mathcal{G}^{A\,\mu\nu}+{\bar \Q_i}(i\slashed{D}-m_{\Q_i})\Q_i+(y_{ij} H \Q_i \Q_j+
\tilde y_{ij} H^* \Q_i \Q_j +\hbox{h.c.})
\ee 
where  $\mathcal{G}_{\mu\nu}^{A}$ is the field-strength for the DC interactions. A topological term for the DC sector can be added,
but it will not play an important role in the present paper. 
When Yukawa couplings are allowed by the gauge quantum numbers, 
two independent couplings  $y$ and $\tilde{y}$ exist for left and right chiralities
of the vector-like fermions, breaking in general parity P and CP.
The addition of new vector-like fermions charged under a dark gauge interaction
maintains the successes of the SM
for what concerns flavor and precision observables. As a consequence, the new physics can lie  around the weak
scale with no tension with experimental bounds, yet accessible to DM and  collider experiments.

\medskip

The renormalizable theories considered here enjoy
accidental symmetries (dark baryon number, species number and generalisations of $G$-parity~\cite{hill})
that lead to stability of particles that are therefore good DM candidates, if safe from decay by dimension five operators
of the form $(\bar \Q_i \Q_j )(H^\dagger H)$.
We focus on the simplest and more robust possibility: DM as the lightest dark-baryon,
made of $\Q^{\ndc}$.
In fact, taking a GUT or a Planck scale as UV cut-off for our model, the approximate dark baryon number conservation 
is typically sufficient to guarantee stability over cosmological time scales. 

Stability of the $\Q^{\ndc}$ dark baryon can remain preserved up to dimension-6 operators
in the presence of extra states charged under $G_{\rm DC}$, 
provided that they have quantum numbers different from $\Q$.
Their thermal relic abundance would be sub-leading, if they are much lighter than $\Q$.
For example, sticking to fundamentals of $G_{\rm DC}$,
 the $\Q\to-\Q$ symmetry remains preserved in the presence of a
 dark scalar ${\cal S}$, as long as fermion singlets $\nu_R$ 
 and the consequent $\Q{\cal S}^* \nu_R$ operators  are absent.

Choices of the gauge quantum numbers that lead to acceptable DM candidates have been presented in the literature~\cite{Antipin:2015xia}. 
We will adopt the simplest and most successful models.

The new point of this paper is that we will study the phenomenology of such models
assuming that the constituent dark quarks have masses $m_\Q$ larger than the confinement scale of the dark gauge interactions
 \be
\LDC\approx \mQ\exp\left[-\frac{6\pi}{11\,C_2(G)\,\adc(\mQ)}\right]
\label{eq:LDC}
\ee
where $C_2(\SU(N))=N$, 
$C_2(\SO(N))=2(N-2)$\footnote{This differs from~\cite{Antipin:2015xia}
because we use a different convention for the normalization of $\adc$, reflected
by the different  index $T=2$ for the vector of $\SO(N)$, see table~\ref{tab:Ccolor}.
The present normalization satisfies $\alpha_{\SO(3)} = \alpha_{\SU(2)}$.}
and $\adc(\mQ)$ is the value of the coupling at the scale of the lightest dark quark.
The temperature at which the dark confinement phase transition occurs roughly is $\LDC$.



This scenario presents qualitatively novel aspects.
Freeze-out of DM constituents $\Q$ occurs at the scale $\mQ/25$ (or larger if there is a dark baryonic asymmetry~\cite{1706.02722}).
At lower temperatures, $\Q$ forms an interacting fluid with dark gluons and possibly with some SM vectors.
DM baryons only form in a second `darkogenesys' stage at a lower temperature, around the dark confinement scale $\LDC$
which could be as light as $100\MeV$. For dark quark masses in the TeV range this translates into
\begin{equation}\label{eq:adcmQ}
\adc(\mQ) \approx \frac{6\pi}{11 C_2(G) \ln m_\Q/\LDC}\approx 0.06  \frac{ {3}/ {C_2(G)}}{\ln  m_\Q/(10^{4}\LDC)}.
\end{equation}
During a first order phase transition, a fraction of the dark quarks manage to form dark baryons,
which remain as DM,
and the remaining fraction annihilates into dark glue-balls, which later decay into SM particles.

\subsection{Models}\label{models}

In the heavy quark regime, $\mQ\gg  \LDC$, the dark baryon mass is roughly the sum of the constituent masses.
Then, mixing between baryons made of different species is negligible as long as
their mass splitting is larger than the binding energy
\beq \left| m_{\mathcal{Q}_1}-m_{\mathcal{Q}_2} \right| > \max (\LDC, \adc^2 m_{\mathcal{Q}_1}).\eeq
We will assume that this is the case, such that DM is made of the lightest specie of dark quarks.
Then, different gauge quantum numbers of $\Q$ give different models.
They fall into two main categories: either $\Q$ is a neutral singlet $N$ under the SM gauge group,
or it is charged.
In the first case the  DM candidate is $\Q^{\ndc}$: a  dark-baryon with spin $\ndc/2$, singlet under the SM.
In the second case DM has lower spin.

Let us discuss  more in detail theories with charged $\Q$.

In theories with dark gauge group $G_{\rm DC}=\SU(\ndc)$
candidates with non-vanishing hypercharge are excluded by direct DM searches,
so that  a successful DM candidate is obtained  if the lightest dark quark is a triplet $V$ under $\SU(2)_L$,  neutral under $\SU(3)_c\otimes{\rm U}(1)_Y$\footnote{An exception can be provided by models with degenerate dark quarks~\cite{1503.04203}.}.
Avoiding sub-Planckian Landau poles for $\SU(2)_L$ fixes $N_{\rm DC}=3$.  

\medskip

The situation is different in theories with dark gauge group $G_{\rm DC}=\SO(\ndc)$:
since its  vectorial representation is real, the lightest dark baryon is a real particle, fermion or boson.
Real particles cannot have a vector coupling to a spin-1 particle, so dark quarks
with non-vanishing hypercharge are allowed as long as a small coupling with the Higgs splits the
two degenerate real states.
Acceptable DM candidates are obtained again for $\Q = V$,
but also for $\Q=L\oplus N\oplus\ldots$ 
or $\Q=L\oplus V\oplus\ldots\;$, 
where the lightest dark quark $L$ 
has the same gauge quantum numbers of a lepton doublet, such that Yukawa couplings to the Higgs are allowed.
Such models can give rise to inelastic dark matter phenomenology~\cite{TuckerSmith:2001hy}.

\begin{figure}[!t]
\begin{center}
\includegraphics[width=.65\textwidth]{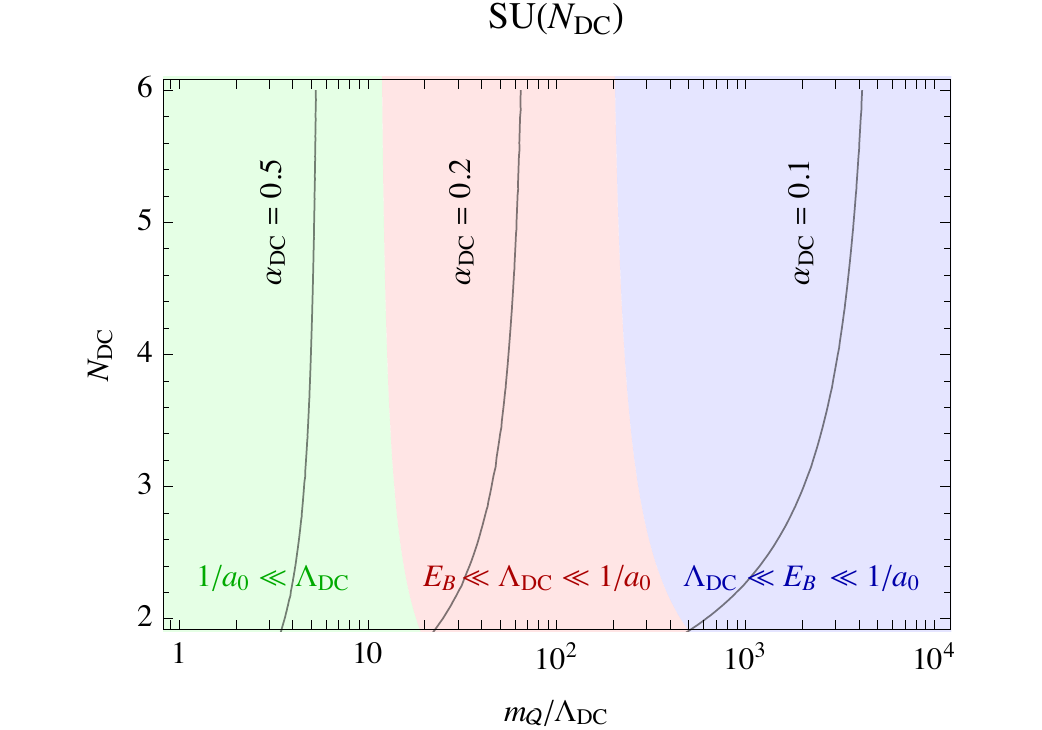}
\caption{\em Qualitatively different regions described in the text
as function of the mass hierarchy $m_\Q/\LDC$ and of $\ndc$,
superimposed to a contour plot of $\adc$ renormalized at $m_\Q$.
We assumed a $\SU(\ndc)$ gauge group;  similar results hold for $\SO(\ndc)$.
\label{fig:Coulombic}}
\end{center}
\end{figure}

\section{The bound states}\label{bound}
Dark gluons form  dark glue-balls (DG),
with mass $\mDG \approx 7 \LDC$.
Dark quarks bind into dark mesons and dark baryons.
In the Coulombic regime the size of dark quark bound states is set by the Bohr radius,
$a_0\sim 1/(\adc \mQ)$ with binding energy  $E_B \sim\adc^2 \mQ $.
We can distinguish three different regimes,  depending on the relative ordering of $1/a_0$, $E_B < 1/a_0$ and $\LDC$: 
\begin{itemize}
\color{blue}
\item[A)] \color{black}
If $\LDC \ll  E_B \ll 1/a_0$:
confinement gives small corrections and bound states are well described by Coulombic potentials.
This region roughly corresponds to $\adc\circa{<}0.1$ and
$m_\Q \circa{>}10^3\LDC$ and is plotted in blue in fig.\fig{Coulombic} for
a $\SU(\ndc)$ group.

\color{red}
\item[B)] \color{black}
If $E_B  \circa{<} \LDC \circa{<} 1/a_0$ dark baryons form 
at temperatures around the confinement scale in excited states, that later try to decay into
lowest lying Coulombian bound states \cite{1606.00159}. This region is plotted in red in fig.\fig{Coulombic} and roughy corresponds to 
$\adc\sim 0.2$ and
$m_\Q \sim100\LDC$.

\color{verdes}
\item[C)] \color{black}
If $ 1/a_0 \ll \LDC$  bound states are similarly to quarkonium   in  QCD,
dominated by confinement phenomena.
This region is plotted in green in fig.\fig{Coulombic}
 and roughy corresponds to 
$\adc\circa{>} 0.4$ and
$m_\Q \circa{<}10\LDC$.

\end{itemize}

\subsection{Dark glue-balls}

\begin{figure}[t] 
\begin{center}
  \includegraphics[scale=0.6]{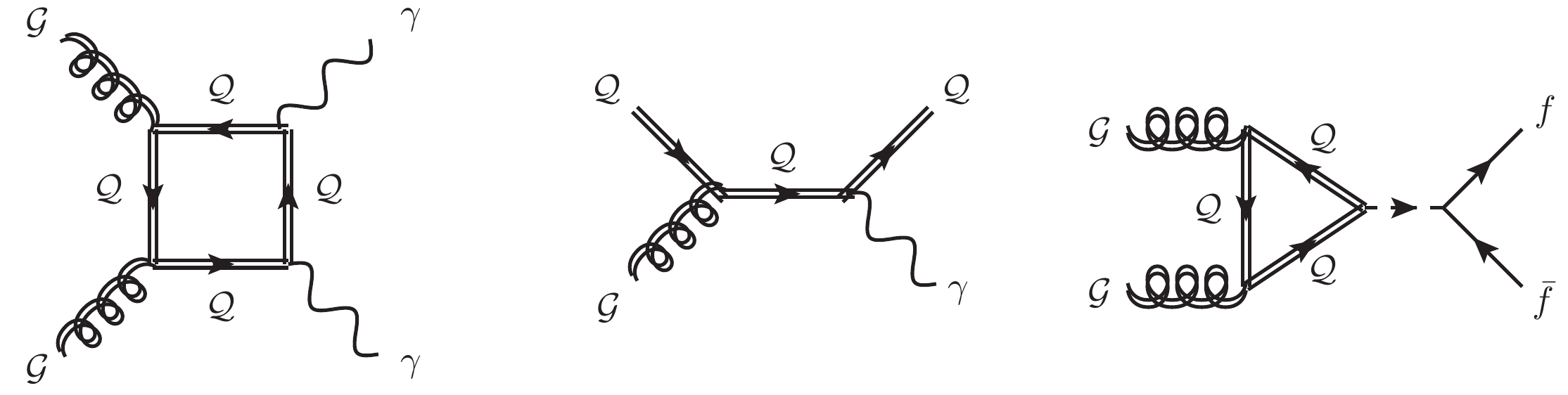}
\end{center}
\caption{\emph{Leading processes describing interactions between the SM and the dark gluons.}}
\label{fig:tcballdecay}
\end{figure}

Under our assumptions, the lightest bound state in the dark sector are dark glue-balls (DG),
with quantum numbers $J^{PC}=0^{++}$ and mass $\mDG \approx 7 \LDC$ ~\cite{9901004}, 
which can be much lighter than the DM mass, $\ndc m_\Q$.
Interactions of dark gluons with the SM are induced by loops of dark quarks (possibly DM itself) charged under the SM sector as in fig.~\ref{fig:tcballdecay}.
Assuming dark quarks with electro-weak charges we estimate the lifetime of the lightest  $0^{++}$ glue-ball as (see section~\ref{DG2} and~\cite{1704.02347})  
\begin{equation}
\tauDG\sim \left\{\begin{array}{ll}
\displaystyle
10\sec \bigg( \dfrac {10 \GeV}{\mDG }\bigg)^9 \bigg( \dfrac {\mQ}{\TeV}\bigg)^8 & \hbox{DG}\to \gamma\gamma\\[4mm]
\displaystyle
10^{-3}\sec\,\left(\frac {0.1}{y}\right)^4 \bigg(\frac{m_b}{m_q}\bigg)^{2}
\bigg( \dfrac {10 \GeV}{\mDG }\bigg)^7 \bigg( \dfrac {\mQ}{\TeV}\bigg)^4& \hbox{DG}\to q \bar q,\hbox{ if }
\mDG >2m_q\\[4mm]
\displaystyle
10^{-16}\sec\,\left(\frac {0.1}{y}\right)^4 \bigg( \dfrac {500 \GeV}{\mDG }\bigg)^5 \bigg( \dfrac {\mQ}{\TeV}\bigg)^4 & \hbox{${\rm DG}\to hh$,
if $\mDG >2M_h$},
\end{array}\right.
\label{eq:Glifetime}
\end{equation} 
where $m_q$ is the mass of the SM quarks. A  smaller life-time arises in the presence of extra light states charged under $G_{\rm DC}$, for example a dark color scalar coupled to the SM through the Higgs portal.
The glue-ball lifetime can vary from cosmological to microscopic values.
As we will see, cosmological constraints generically imply\footnote{We do not consider cosmologically stable glue-balls as DM candidates because their 
thermal abundance is too large if the dark sector was in thermal equilibrium with the SM.}
\begin{equation}
\tauDG + t_{\LDC} \circa{<} 1\sec
\end{equation}
where  $t_{\LDC}\sim M_{\rm Pl}/g_*^{1/2}\LDC^2$ is the cosmological time at which dark confinement occurs.

\begin{table}[t]
\begin{center}
\begin{tabular}{c|c|ccc}
\rowcolor[HTML]{C0C0C0} 
$G_{\rm DC}$ & Representation $R$ & Dimension $d$ & Index {\color[HTML]{000000} $T$} & Casimir {\color[HTML]{000000} $C$} \\\toprule
& fundamental & $N$  & $\sfrac{1}{2}$ & $\sfrac{(N^2-1)}{2N}$         \\
\multirow{-2}{*}{$\SU(N)$} & ${\rm adjoint}$ & $N^2-1$ & $N$ & $N$\\\midrule
& fundamental & $N$ & 2 & $N-1$\\
\multirow{-2}{*}{$\SO(N)$} & ${\rm adjoint}$ & $\sfrac{N(N-1)}{2}$ & $2N-4$ & $2N-4$     \\     \midrule           
\end{tabular}\quad\quad

\end{center}
\caption{\em The dimension, the index $T$ and the quadratic Casimir $C$ of fundamental and adjoint $\SU(N)$ and $\SO(N)$ representations. 
\label{tab:Ccolor}}
\end{table}

\subsection{Dark mesons}
Dark confinement implies that physical states at zero temperature are singlets of dark color: mesons and baryons. 
Assuming that dark quarks fill a representation $R = (R_{\rm DC}, R_{\rm SM})$
of the dark gauge group times the SM gauge group,
the non-relativistic interaction between a $\Q$ and a $\bar \Q$ is
a Coulomb/Yukawa potential
mediated by dark vectors and by SM vectors.
For a two-body state in the representation $J_{\rm DC}\in R_{\rm DC}\otimes \bar R_{\rm DC}$
of $G_{\rm DC}$ and
$J_{\rm SM}\in R_{\rm SM}\otimes \bar R_{\rm SM}$ of $G_{\rm SM}$
 the Coulombic potential is
\be\label{eq:coulombpot}
V=-\frac{\adc\lambda_{\rm DC}+\alpha_{\rm SM}\lambda_{\rm SM}}{r}\equiv-\frac{\alpha_{\text{eff}}}{r},\qquad\lambda_{J}=\frac{C_{R_J}+C_{\bar R_J}-C_J}{2},\qquad 
\ee 
where $C_{R_J}$ are the quadratic Casimirs, see table \ref{tab:Ccolor}.
In the Coulombic regime the size of dark quark bound states is given by the Bohr radius,
$a_0\sim2/(\alpha_{\textrm{eff}}\mQ)$ while the energy is $E_B \sim \alpha_{\textrm{eff}}^2 \mQ /4$.
For  $\Q\bar{\Q}$ dark meson singlets one finds $ \alpha_{\textrm{eff}}= C_N \adc$.

When $a_0 > \LDC^{-1}$ the effects of confinement cannot be neglected.
The effective potential can be approximated as $V \approx - \alpha_{\textrm{eff}} /r + \LDC^2 r$
so that the bound states are dominated by the Coulombian term when $\LDC^2 a_0^2 < \alpha_{\textrm{eff}}$ 
or equivalently    $\LDC/m_\Q \circa{<}\adc^{3/2}$: the Coulombic approximation does not hold in the green region of
fig.~\ref{fig:Coulombic}.

\subsection{DM  dark baryons}

Under our assumptions DM is the neutral component of dark baryons made of the lightest dark-quark multiplet\footnote{Electro-weak
interactions split the neutral from the charged components of $\SU(2)_L$ multiplets ($\Delta \mQ=\alpha_2 M_W \sin^2( \theta_{\rm W}/2) \approx 165\MeV$ when hypercharge vanishes~\cite{hep-ph/0512090}). In our region of parameters $\mQ\gg \LDC \circa{>}$ GeV the mass splitting is always smaller than the binding energy of the baryons so that we can work in an approximate $\SU(2)_L$ invariant formalism.}. 
The lightest dark baryons are the $s$-wave bound states with minimal spin (altought extra spin gives a small extra mass, unlike in QCD).

If the lightest dark quark is a SM singlet, $\Q=N$, the lightest dark baryon has a symmetric spin wave-function, so that its spin is $\ndc/2$. 
If instead $\Q$ has a multiplicity $N_F$ the lightest baryons fills the following representations, under both
flavour and spin:
\begin{equation}
\hbox{lightest dark baryon} = \left\{\begin{array}{ll}
{\tiny \Yvcentermath1  \yng(2,1)} & \hbox{for $\ndc =3$}\\
{\tiny \Yvcentermath1  \yng(2,2) } & \hbox{for $\ndc =4$}\\
{\tiny \Yvcentermath1\yng(3,2) } & \hbox{for $\ndc =5$} 
\end{array}\right.
\label{youngtableauxlight}
\end{equation}
so that their spin is either 0 (for $\ndc $ even) or 1/2 ($\ndc $ odd). 
For example in the model where $\Q=V$ (a $\SU(2)_L$ triplet)
and $G_{\rm DC}=\SU(\ndc)$, the lighter dark baryons are triplets under $\SU(2)_L$ for $\ndc$ odd and singlets for $\ndc$ even. 

\medskip

The binding energy of dark baryons can be computed precisely using variational techniques.
Let us consider a more general system made of $n\le\ndc$ SM singlets dark quarks $N$ in the anti-symmetric dark-color configuration.
In the non-relativistic limit the Hamiltonian is
\beq H =K +V,\qquad 
K=\sum_{i=1}^{n} \frac{p_i^2}{2m_\Q},\qquad
V=- \frac{C_N \adc }{ N_{\rm DC}-1}  \sum_{i<j}^{n} \frac{1}{r_{ij}}\eeq
where $\mb{r}_i$ is the position of dark-quark $i$ and $r_{ij} = |\mb{r}_i - \mb{r}_j|$.
It is convenient to rewrite  $H$ in terms of 
the center-of-mass coordinate 
$\mb{X} = \frac{1}{n}\sum_{i=1}^n \mb{r}_i$,
of the associated canonical momentum  $\mb{P} = \sum_{i=1}^n \mb{p}_i$,
and of the distances
$\mb{\delta}_i = \mb{r}_i - \mb{r}_n$ 
with associated canonical momenta $\mb{\pi}_i =\mb{p}_i -  \mb{P}/n$ for
$i=1,\ldots n-1$ 
The kinetic energy becomes
\beq  K =
K_{\rm CM} + \frac{1}{m_\Q}\sum_{i\ge j}^{n-1}\mb{\pi}_i \cdot\mb{\pi}_j
\eeq
where $K_{\rm CM} = \sfrac{P^2}{2n m_\Q  }$.
We compute the binding energy of the lightest baryons using  the variational method with  
trial wave-functions for the dark-baryon state $|\B\rangle$  containing one parameter $k$ with dimensions of inverse length.
Defining $\langle X\rangle =  {\langle \B |X | \B\rangle}/{\langle \B | \B\rangle}$
we use  $\mb{\pi}_i = - i\partial/\partial\mb{\delta}_i$ 
and parameterize  $\langle 1/r_{ij}\rangle = C_V  k$  and
$\langle K - K_{\rm CM}\rangle = n C_K k^2/2m_\Q$
such that
\beq \langle H - K_{\rm CM} \rangle= n C_K 
 \frac{k^2}{2m_\Q}- C_V k\, \frac{n(n-1)}{2} \frac{C_N \adc}{\ndc-1}.
\eeq
Maximising with respect to $k$ gives the binding energy
\beq \label{eq:EBbaryon}
E_B^{\Q^n} =C_E C_N^2 \alpha_{\rm DC}^2 m_{\mathcal Q}\times \frac{(n-1)^2}{(\ndc-1)^2}\qquad
C_E =  \frac {nC_V^2} {8C_K}\eeq
where the last factor equals 1 for dark baryons with $n=\ndc$.

\begin{table}
$$\begin{array}{|c|ccc|ccc|ccc|} \hline
& \multicolumn{9}{|c|}{\hbox{Trial dark-baryon wave-function $\psi_\B(\mb{r}_1,\ldots \mb{r}_n) $}}\\ 
 &\multicolumn{3}{|c|}{ \exp(-k\sum_{i<j}^n r_{ij}) }& 
 \multicolumn{3}{|c|}{\sum_{i=1}^n \exp(-k\sum_{j=1}^n r_{ij})} &
 \multicolumn{3}{|c|}{\exp(-k\sum_{i=1}^n r_{i}) } \\
 \hline
n & C_V & C_K & C_E & C_V & ~~C_K~~ & C_E  & C_V & C_K & C_E \\ \hline
2 & 1  &1 & 0.25 &  1 & 1 & 0.25     &  5/8 &1 & 0.10\\
3 & 1.43 &2.8 & 0.27  &   0.92 & 1.22 & 0.26   & 5/8 & 1 &0.14 \\
4 & 1.7 &5&0.28& 0.88 & 1.3 & 0.29 &      5/8 & 1 &0.19 \\
5 & &&&   0.85&1.4&0.33   & 5/8 & 1 & 0.24 \\
6 & &&&  \sim 0.8&\sim 1.2&\sim 0.4   & 5/8 & 1 & 0.29 \\ \hline
\end{array}$$
\caption{\label{tab:EB}\em Binding energies of anti-symmetric bound states made of $n$ dark-quarks with mass $m_\Q$
with a non-abelian Coulombian potential.
We use  the variational method and assume three different trial wave functions.
The coefficients $C_{V,K,E}$ are defined in eq.\eq{EBbaryon}.
In particular,
for $n=\ndc$ the bound states are dark baryons, and $E_B = C_E (C_N \adc)^2 m_\Q$.
}
\end{table}

\medskip
 
Table~\ref{tab:EB} shows the resulting coefficients for three different trial wave-functions.
For $n=2$ we reproduce the Coulombian binding energy.
For $n=3$ and gauge group $\SU(3)$ we reproduce the QCD result,
$E_B^{\Q\Q\Q} \approx 0.46  \adc^2 \mQ$~\cite{hep-ph/0607290} (see also~\cite{1008.3154}).
Numerical integration becomes increasingly difficult for higher $n$.
 
The first two trial wave-functions depend only on relative distances $r_{ij}$ and give similar results
for the binding energy (the biggest result is the best approximation).
The third wave-function
$\psi_\B = (k/\pi)^{n/2} \exp(-k\sum_{i=1}^n r_{i})$, considered in~\cite{1003.4729} for $G=\SU(\ndc)$,
depends on absolute coordinates $r_i$, 
such that  the center-of-mass kinetic energy is not subtracted:
it leads to $C_V =5/8$ and $C_K=1$ for any $n$
(we find order one factors that differ from the analogous computation in~\cite{1003.4729}),
and the resulting binding energy can be a reasonable approximation at large $n$.

As the numerical computation becomes more difficult for large $\ndc$, it useful to complement it with the following approximation.
The binding energy of dark baryons can be semi-quantitatively understood by building them recursively adding 
dark quarks to a bound state.
For $G_{\rm DC}=\SU(3)$ the baryon can be thought as a stable di-quark bound to a quark. 
Treating the di-quark as elementary we can construct a color singlet baryon adding the third quark. 
Summing up the binding energies of $\Q\Q$ and $\Q\Q+\Q$ one finds $E_B\sim 0.7 \adc^2 \mQ$ not far from the correct value $E_B\sim 0.45 \adc^2 \mQ$.
Because the gauge wave-function of di-quarks is anti-symmetric, the spin of $s$-wave bound states is 1 for a symmetric flavor wave-function 
and 0 for an anti-symmetric wave-function. 
Generalising this argument to $\ndc $ quarks one finds a Bohr radius $a_0^{-1} \approx \adc \ndc  \mQ$ and a binding energy  $E_B \approx \adc^2 N_{\rm DC}^3 \mQ$ in agreement with \cite{Witten:1979kh}.\footnote{The binding energy of $n-1$ antisymmetric dark quarks with an extra dark quarks is
$
E_B^{n-1,1}= \frac12 \lambda_{n-1,1,n}^2 \adc^2  {\mu_{n-1,1}}$
where $\mu_{n_1,n_2}= n_1 n_2/(n_1+n_2) \mQ$ is
the reduced mass and $\lambda_{n_1,n_2, n_3}=(C_{n_1}+C_{n_2}-C_{n_3})/2$.
The quadratic Casimir of the $n$-index antisymmetric tensor  of $\SU(\ndc)$ is $C_n= \frac12{n(\ndc -n)}(1+  1 /{\ndc })$.
 The total binding energy of a singlet made of the anti-symmetric combination of $n = N_{\rm DC}$ dark quarks  is then
\begin{equation}
E_B^{\Q^n} \approx \sum_{n=2}^{\ndc }E_B^{n-1,1} \approx \frac {\ndc^2(\ndc -1)}{24}\adc^2\mQ.
\end{equation}}

\subsection{Annihilations of DM dark baryons}\label{DMBann}
Annihilations of DM dark baryons are relevant for computing their cosmological thermal abundance
(section~\ref{relic}) and for indirect detection signals (section~\ref{indirect}).

The cross section for annihilation of dark baryons $\B$ with dark anti-baryons $\bar \B$
 receives a contribution of particle-physics size, due to
perturbative annihilation of constituents, $\sigma_{\B\bar \B} v_{\rm rel} \sim \pi \adc^2/m_\Q^2$.
A bigger contribution arises at scattering energies smaller than the binding energy:
the long-range Coulomb-like force inside baryons can distort the orbits of the constituent quarks
such that two overlapping baryons can recombine into mesons.
Despite the negligible energy transfer this rearrangement has a large effect, because the $\Q\bar\Q$ into mesons 
later annihilate, such that mesons decay.\footnote{This phenomenon is
somewhat analogous to the annihilation of hydrogen ($ep$) with anti-hydrogen ($\bar e\bar p$), that can recombine as $(ep)+(\bar e \bar p) \to (e\bar e)+ (p\bar p)$ followed by the $e\bar e$ and $p \bar p$ annihilation processes.
Recombination is energetically favourable because the two heavier protons can form a deep bound state.
The rearrangement cross section is of  atomic size,   $\sigma v_{\rm rel}\sim \sqrt{m_e/m_H}\pi\alpha_{\rm em} a_0^2$
for $m_H v_{\rm rel}^2 < m_e\alpha_{\rm em}^2 $~\cite{Morgan:1970yz,Morgan:1973zz,froelich,Jonsell:2000wb}.}
Such recombination can take place efficiently only if $v_{\rm rel} \circa{<} \adc$:
classically this corresponds to the condition that the relative velocity is not much larger than the
orbital velocity;
quantistically to the condition that the wave-length of the incoming 
particles is larger than the size of the bound states.
At larger energy one has partonic scatterings among constituents, with the smaller cross section discussed above.

The dominant recombination, if allowed kinematically, arises when a dark baryon $\Q^{\ndc}$ and
a dark anti-baryon $\bar\Q^{\ndc}$
emit one $\Q\bar\Q$ dark meson, leaving a dark baryonium bound state made of $\ndc-1$ dark quarks $\Q$
and $\ndc-1$ anti-quarks $\bar\Q$:
\begin{equation}
(\Q^{\ndc})+(\bar \Q^{\ndc}) \to (\Q\bar\Q) +  (\Q^{\ndc-1})(\bar \Q^{\ndc-1}).
\label{eq:1mesonrearrengement}
\end{equation}
Rearrangements into several mesons, such as  $(\Q^{\ndc})+(\bar \Q^{\ndc}) \to (\Q\bar\Q)^{\ndc}$, 
is suppressed at large $\ndc$~\cite{Witten:1979kh}. 

Assuming an estimate similar to the hydrogen-anti-hydrogen result,
the cross-section relevant for indirect detection and at late times during the freeze-out is
\begin{equation}
\sigma_{\B\bar \B} \sim \frac{\pi \, R_{\B}^2}{\sqrt{E_\text{kin}/E_B}}\qquad \Rightarrow
\qquad \sigma_{\B\bar \B} v_{\rm rel} \ \sim \frac 1 {\sqrt{N_{\rm DC}}C_N \adc} \frac {\pi}{ \mQ^2} 
\label{eq:indirectxsec}
\end{equation}
which vastly exceeds the annihilation cross sections among dark-quark constituents, $\sigma_{\Q\bar\Q} v_{\rm rel} \sim \sfrac {\pi \adc^2}{\mQ^2}$. Heuristically the large cross-section can be understood as follows:
when the baryon-anti-baryon overlap a quark anti-quark-pair becomes unbound and can form a meson. 
For low enough velocities this process happens with probability of order one leading to an almost geometric cross-section.
Additionally we consider thermal correction to the Bohr radius, which can become important during the freeze-out process \cite{1606.00159}. A more precise value of  $\sigma_{\B\bar \B}$ needs a dedicated non-relativistic quantum mechanical computation.

Next, we can check which rearrangements are kinematically allowed.
Considering, for example,  $G_{\rm DC} =\SU(3)$ 
($C_N = \frac43$)  or $\SO(3)$ ($C_N = 2$)
 we have the following binding energies:
\begin{itemize}
\item The binding energy of a $\Q\bar\Q$ singlet meson is $E_B^{\Q\bar\Q}=\frac14 C_N^2 \adc^2 \mQ$, see the discussion around eq.\eq{coulombpot}.
\item The binding energy of a $\Q\Q\Q$ baryon is $E_B^{\Q\Q\Q}\approx 0.26 C_N^2\adc^2 m_\Q $, see eq.\eq{EBbaryon}.
\item The binding energy of a $\Q\Q$ di-quark state is $ E_B^{\Q\Q}=\frac14 E_B^{\Q\bar\Q}$, see eq.\eq{EBbaryon}. 
\end{itemize}
The rearrangement into 3 mesons is kinematically allowed, given that the energy difference is positive:
$\Delta E_B= 3 E_B^{\Q\bar\Q}- 2 E_B^{\Q\Q\Q}\approx 0.23C_N^2 \adc^2\mQ$.

The dominant process in eq.~(\ref{eq:1mesonrearrengement}) seems also allowed, in view of
\beq \Delta E_B = E_B^{\Q\bar\Q}+ E_B^{\Q\Q\bar\Q\bar\Q}-2 E_B^{\Q\Q\Q}\approx
(1+2)E_B^{\Q\bar\Q} +2 E_B^{\Q\Q} -2 E_B^{\Q\Q\Q}=0.35 C_N^2 \adc^2\mQ > 0\eeq
where we estimated the binding energy of $\Q\Q\bar\Q\bar\Q$ as the one of $\Q$-$\Q$ and of $\bar\Q$-$\bar\Q$,
plus the $(\Q\Q)$-$(\bar\Q\bar\Q)$ binding energy approximated as $2 E_B^{\Q\bar \Q}$, where the factor of 2 accounts for 
the reduced mass.

If the dark baryons $\B$ are not in the Coulombic regime, they can be approximated as heavy dark quarks kept together by flux tubes
which give a confining linear potential $V \sim \LDC^2 r$.
The recombination cross section then is geometric, $\sigma_{\B \bar \B} \sim \pi R^2$, at any scattering energy~\cite{hep-ph/0611322,0712.2681,1112.0860}.
Indeed this is the cross section for crossing of two flux tubes with length $\approx R$;
lattice simulations suggest that the probability of reconnection is close to one
(a similar process takes place in string theory, where the reconnection probability can be suppressed by the string coupling~\cite{hep-th/0412244}).

\begin{figure}[t]
\begin{center}
$$\includegraphics[width=.6\textwidth]{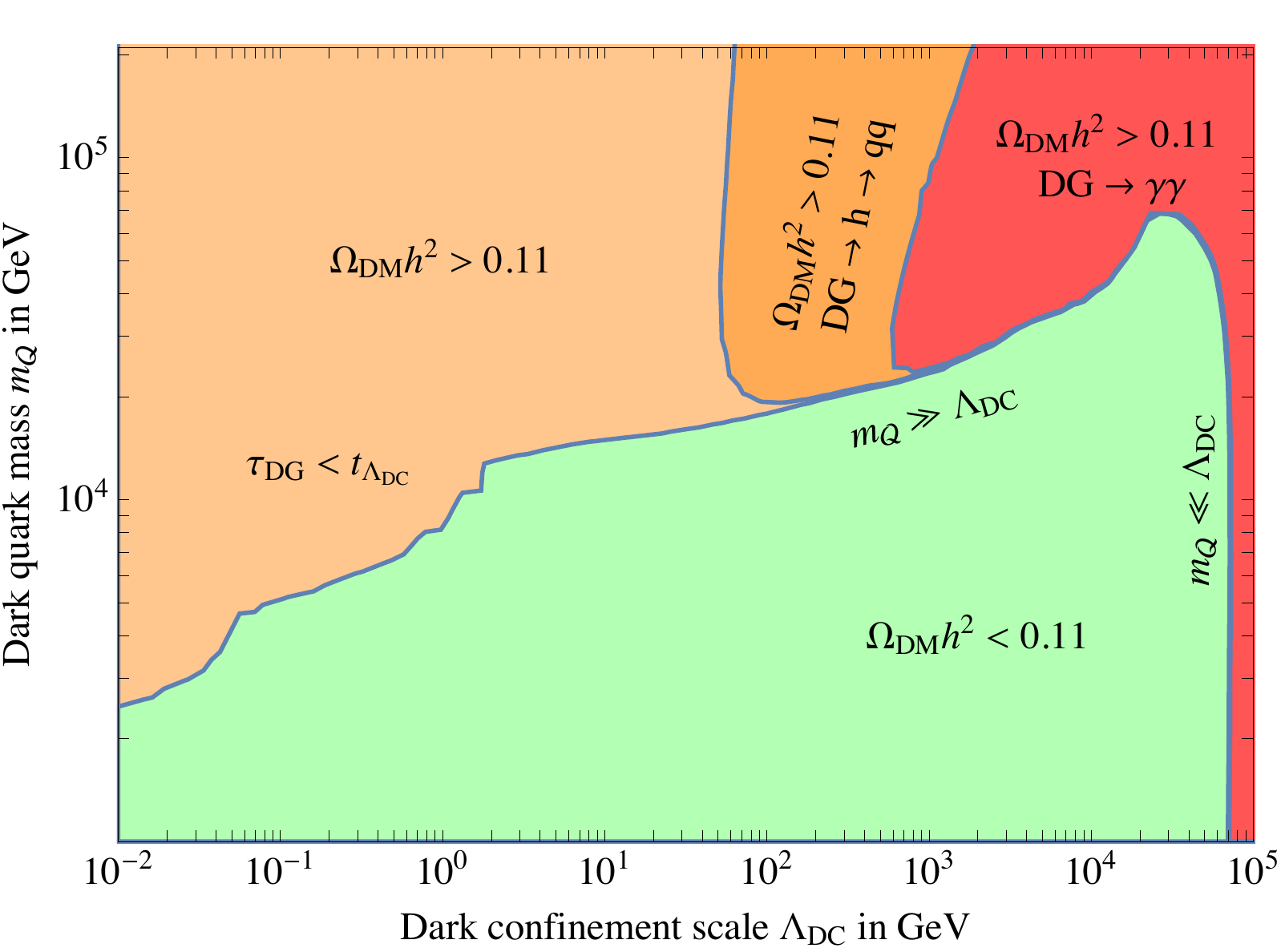}$$
\caption{\em Qualitative dependence of the DM relic abundance as function
of $\LDC$ and of $m_\Q$: the cosmological value is reproduced along the boundary between the green
and red regions. For $\LDC \ll m_Q\approx 100\TeV$ we recover the results of~\cite{Antipin:2015xia}.
A lighter $m_\Q$ is allowed if instead $m_\Q\gg \LDC$, in view of the perturbative value of $\adc$ at freeze-out.
However, if the glue-ball lifetime $\tauDG$ is too long, glue-ball decays can wash-out the DM density.
We consider 3 different scenarios: decay due to heavy states charged under the SM;
decay due Yukawa couplings to the Higgs with $y \approx 0.2$,
and a shorter life-time, possible due to existence of a light scalar.
\label{fig:DMabundance} }
\end{center}
\end{figure}

\section{DM relic abundance}\label{relic}
We here study the thermal relic DM abundance, assuming a vanishing or negligible dark-baryon asymmetry.
No such asymmetry can exist in $\SO(\ndc)$ models (because baryons are 
real particles), while generating an asymmetry in $\SU(\ndc)$ models requires
substantially more complicated constructions~\cite{1706.02722}.
We need to distinguish two qualitatively different scenarios:
\begin{itemize}
\item{\em Dark color confines before freeze out}, i.e.\ $\LDC\circa{>} \mQ/25$:
dark baryons  form before freeze-out, but their kinetic energy at freeze-out is large
relative to their potential energy, so that the annihilation cross section is the one among
constituents, $\sigma_{\Q\bar \Q} v_{\rm rel} \sim \pi \adc^2/\mQ^2$, smaller than
the cross section in the limit $\LDC \gg \mQ$ considered in previous works~\cite{Antipin:2015xia}.
Thereby  the DM mass suggested by the cosmological abundance is mildly smaller than $M_\B\sim 100\TeV$. 

\item  We focus on the more radical possibility that {\em dark color confines after freeze out}, at $\LDC\ll  \mQ/25$. 
Around  freeze-out at $T \sim \mQ/25$ the dark coupling
$\adc$ is perturbative and dark quarks $\Q$ are free.
They later partially combine into DM baryons at $T \sim \LDC$.
The DM mass suggested by cosmology is smaller than in the previous case.

\end{itemize}
The SM sector and the dark sector are in thermal contact during freeze-out if
$\Q$ is  charged under $G_{\rm SM}$ (for example $\Q$ could be a triplet under $\SU(2)_L$),
or in the presence of a heavier dark quark $\Q'$ charged under the SM, provided that its mass is comparable to $\Q$.
If instead $m_{\Q'}\gg m_\Q$ the two sectors decouple at $T \circa{<} m_{Q'}/25$;
nevertheless they later evolve keeping equal temperatures as long as there are no entropy release takes place.
Otherwise, if the numbers of degrees of freedom $g_{\rm SM}$ or $g_{\rm DC}$ depend on $T$
(this happens in the SM at $T \circa{<}  M_t$),
the  temperatures  become mildly different, satisfying
$g_{\rm SM}(T_{\rm SM}) T_{\rm SM}^3/g_{\rm SM}(T_{\rm dec}) = g_{\rm DC}(T_{\rm DC}) T_{\rm DC}^3/g_{\rm DC}(T_{\rm dec})$.


More importantly, the fraction of the dark energy density which does not contribute to forming DM dark baryons
thermalises into dark glue-balls which decay into SM particles.
These decays only produce a mild entropy release into the SM sector,
 $(T'_{\rm SM}/T_{\rm SM})^3=1+r(g_{\rm DC} T_{\rm DC}^3)/(g_{\rm SM} T_{\rm SM}^3)$ with $r=1$,
provided that $\tauDG < t_{\LDC}$, such that dark glue-balls decay while relativistic.
If instead $\tauDG > t_{\LDC}$,  dark glue-balls can decay while they dominate the energy density,
because the energy density has grown by a factor $r \approx (\tauDG/ t_{\LDC})^{2/3}$ relatively to the SM energy density.
This factor arises as follows.
In a first phase, dark glue-balls are kept in thermal self-equilibrium by `cannibalistic' $3\to 2$ scatterings,
such that conservation of dark entropy $S_{\rm DC} = a^3 (\rho_{\rm DG}+ p_{\rm DG})/T_{\rm DC}$~\cite{1402.3629,1411.3727,1602.00714} implies $\rho_{\rm DG} \stackrel{\propto}{\sim} 1/a^3$,
while $T_{\rm DC}$ evolves only logarithmically with $a$, the scale factor of the universe.
After freeze-out of dark glue-balls, they dilute as non relativistic matter, such that again $\rho_{\rm DG}\propto 1/a^3$.
Given that SM particles are relativistic and dilute as $\rho_{\rm SM} \propto 1/a^4$,
the relative dilution is $\rho_{\rm DG}/\rho_{\rm SM}\propto a$.
The scale factor at the epoch of glue-ball decays 
is estimated from the condition $H(a) \approx \tauDG$.
If glue-balls temporarily dominate the energy budget of the universe,
their decays produce a huge entropy release, washing out the DM abundance as well as the baryon abundance.
The situation is qualitatively illustrated in fig.\fig{DMabundance}.

\begin{figure}[t]
\begin{center}
$$\includegraphics[width=.3\textwidth]{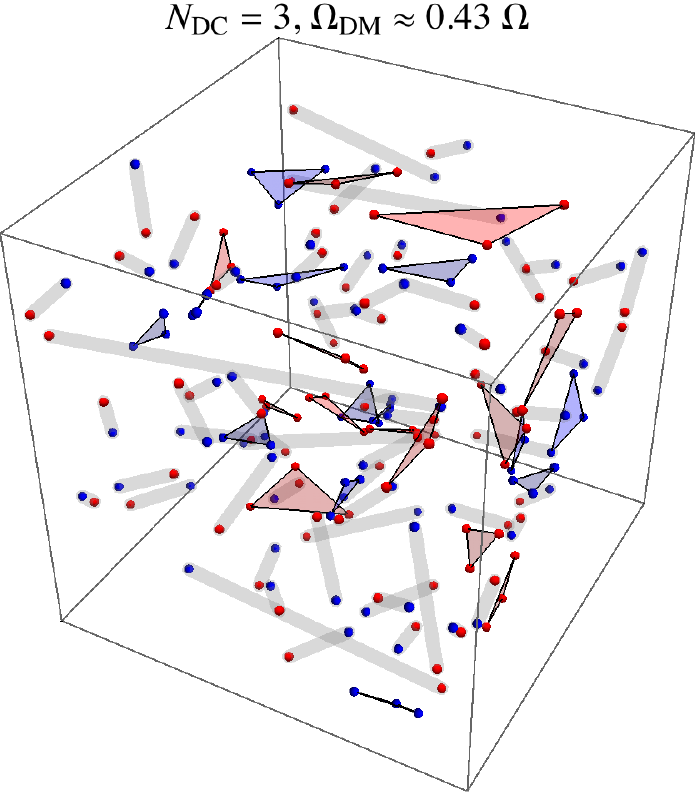}\qquad
\includegraphics[width=.3\textwidth]{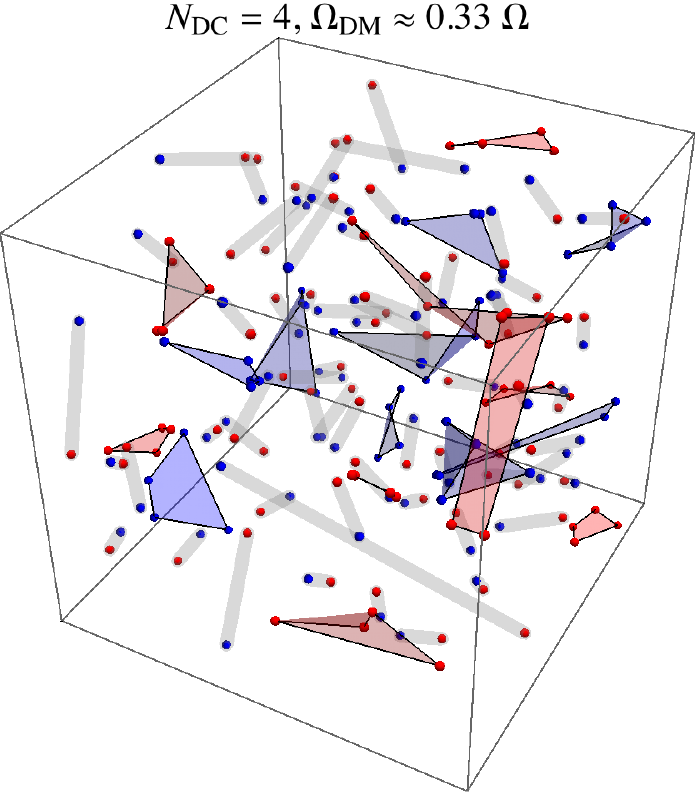}\qquad
\includegraphics[width=.3\textwidth]{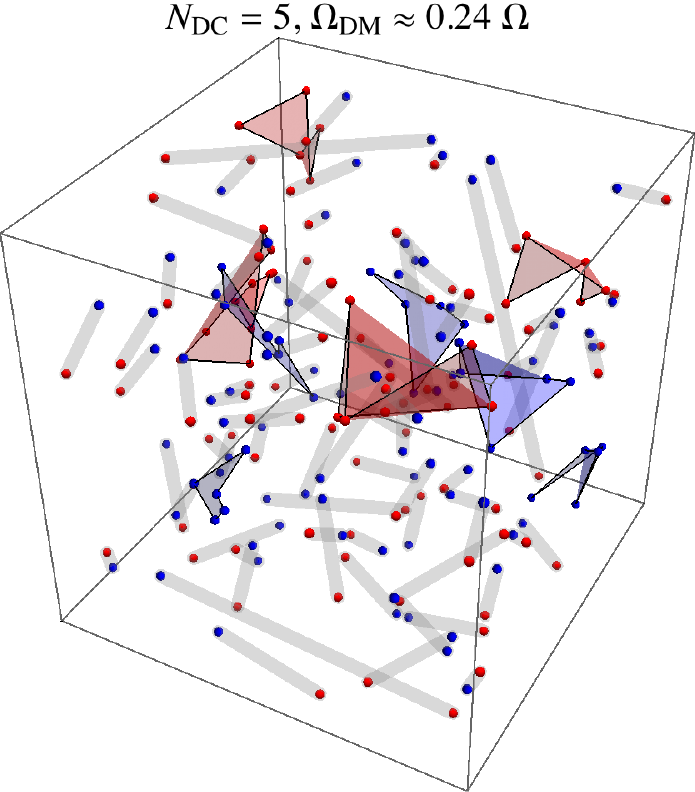}$$
\caption{\em Examples of dark condensation for $\ndc=3$ (left), $4$ (middle) and $5$ (right).
Dark quarks $\Q$ (anti-quarks $\bar\Q$) are denoted as red (blue) dots, placed at random positions.
We assume that each DM particle combines with its dark nearest neighbour, 
forming  either unstable $\Q\bar\Q$ dark mesons (gray lines) 
or stable $\Q^{\ndc}$ dark baryons (red regions)
and $\bar\Q^{\ndc}$ dark anti-baryons (blue regions).\label{fig:DarkCondensation}}
\end{center}
\end{figure}

\subsection{Freeze out of dark quarks and Dark Condensation}
Let us discuss in detail the case where the confinement phase transition takes place after freeze-out,
corresponding to a relatively small $\adc(\mQ)$, see eq.\eq{adcmQ}.

The density of free quarks after freeze-out and before confinement can be computed by solving the coupled Boltzmann equations for the fermions and bound states, described in appendix \ref{sec:AppA}.
Formation of  bound states from dark quarks is a negligible phenomenon 
until the dark gauge coupling is perturbative, given that only a small
amount of dark quarks survived to their freeze-out, as demanded by the observed cosmological DM density. 
Formation of $\ndc \otimes\bar N_{\rm DC}$ and $\ndc\otimes \ndc$ two-body bound states is further suppressed by the fact that it proceeds from a repulsive initial channel given that one dark-gluon must be emitted, in dipole approximation, to release the binding energy.
In appendix~\ref{sec:AppA} we show that only a small fraction of dark quarks gets bound in stable $\ndc\otimes \ndc$ states.

Only when the temperature of the dark sector cools below the dark confinement scale, a dark phase transition happens
(likely first order~\cite{hep-lat/0307017}, leading to potentially observable gravity wave signals), and dark quarks must recombine to form either  dark mesons or dark baryons.
Dark mesons annihilate, heating the plasma of dark glue-balls, which later decay into SM particles.
Only dark baryons survive as DM. Thereby we need to determine the fraction of DM that survives to this phase of dark condensation.

\begin{figure}[t]
\begin{center}
\includegraphics[width=0.6\textwidth]{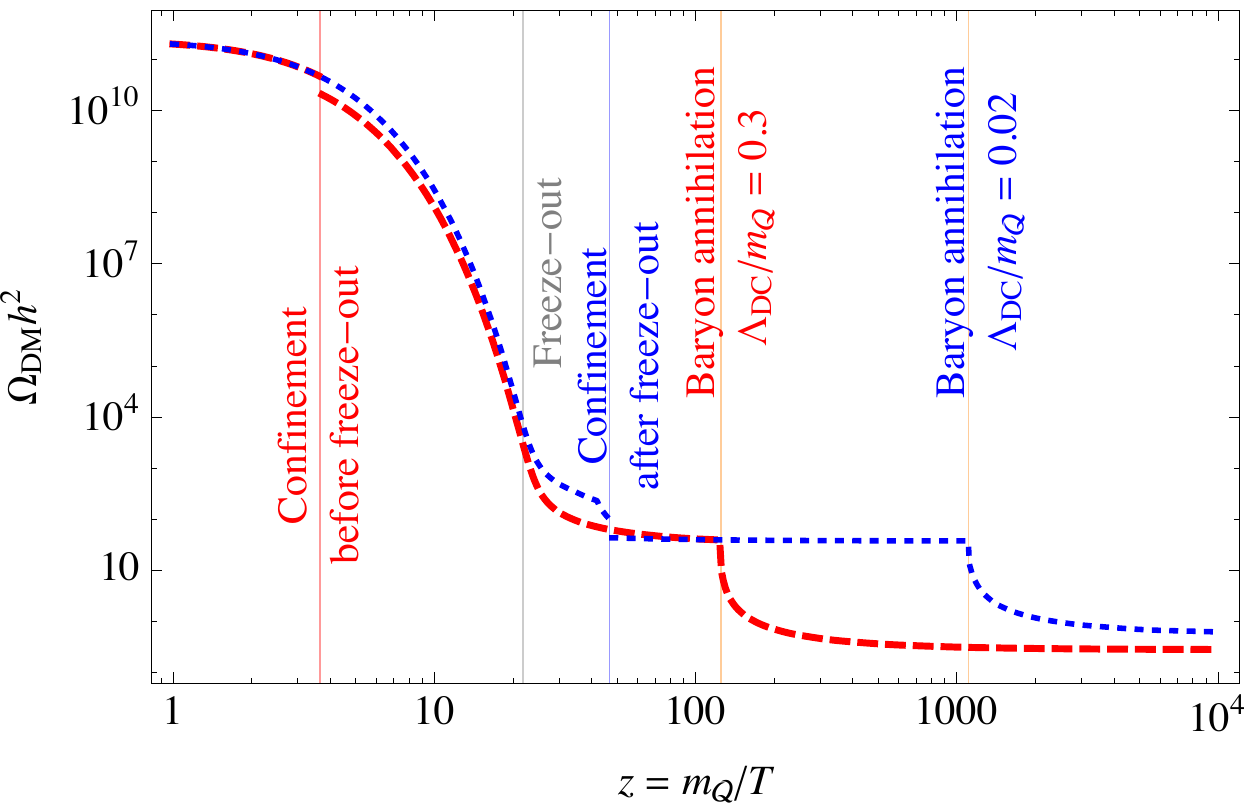}
\caption{\em The freeze-out history of two scenarios is displayed. The red line corresponds to confinement which takes place before freeze-out and the blue line shows the freeze-out which is followed by confinement and condensation. In both scenarios at late times, once the velocity drops below a critical value the constituent annihilation is replaced by a baryonic recombination, which leads to a late stage of dark matter annihilation and an additional depletion of the DM density. \label{fig:RelicRun}}
\end{center}
\end{figure}

Unlike in QCD, dark quarks are much heavier than the confinement scale, 
so that we can neglect the possibility that $\Q\bar \Q$ pairs are created from the vacuum
in order to favour the rearrangement of dark colors~\cite{Webber:1983if,Andersson:1983ia}.
Furthermore, dark quarks form a diluted gas, in the sense that the average
distance $d(\LDC)$ between them is much larger than $1/\LDC$,
\begin{equation}
d(T) \sim \frac 1 {n_\Q(T)^{ 1/3}}\sim \frac 1 T \left(\frac {2\pi T_f}{\mQ}\right)^{\frac 1 2} e^{\sfrac{\mQ}{3 T_f}}
\end{equation}
We are left with a classical combinatorics problem, a geometrical confinement.
Each dark quark is connected to  a string, and the sea of $\Q$ and $\bar\Q$ must recombine into color singlets.
Assuming that a fraction $\wp_\B$ of dark quarks recombines into baryons the required abundance of DM is obtained for
\begin{equation}\label{eq:sigmavann}
\langle\sigma_{\Q\bar\Q} v_{\rm rel} \rangle \approx \frac{\wp_\B}{(23\, {\rm TeV})^2} 
\end{equation}
We  assume in what  follows that $\wp_\B\sim 1$ for small $\ndc \sim 3$. A possible justification goes as follows.
In three dimensions the distance of a dark quark to its nearest neighbour is $0.75$ times smaller than the distance to its next to nearest neighbour, on average.  This suggests that only the nearest neighbours are relevant to the recombination process.
Assuming that each $\Q$ or $\bar\Q$ reconnects  with probability one with  its nearest neighbour,
as illustrated in fig.~\ref{fig:DarkCondensation}, the probability to form a dark baryon is roughly $(1/2)^{\ndc-2}$ smaller than the probability of forming a dark meson.
One than finds
\begin{equation}\label{eq:pB}
\wp_\B\approx \frac{1}{1+2^{\ndc-1}/\ndc} .
\end{equation}
At face value for $\ndc=3$ this gives a baryon fraction 0.4 in agreement with other estimates in the literature. One possible source of error arises from effects of crossing and rearrangement of flux tubes during the recombination process.

So far we assumed no dark-baryon asymmetry.
In $\SU(N)$ models dark baryon number is conserved and, in more complicated models,
a dark-baryon asymmetry could be generated.
Then one would get an extra contribution given by
$\Omega_{\rm DM} = |\Omega_\Q -\Omega_{\rm\bar\Q}|$.

The enhancement in $\sigma_{\B\bar \B}/\sigma_{\Q\bar\Q} \sim 1/\adc^3$ due to recombination,
discussed in section~\ref{DMBann}, 
leads to an extra dilution
of the DM cosmological abundance, see fig.~\ref{fig:RelicRun}.
As the critical cross section relevant for cosmology scales as $\langle \sigma v_\text{rel} \rangle  \propto 1/T$, 
this effect can be relevant provided that  $ m_\Q/\LDC \circa{<} 10^{4}$ is not too large. 

A larger related effect can emerge in the intermediate region B) where $E_B  \circa{<} \LDC \circa{<} 1/a_0$~\cite{1606.00159}.
In this region the lowest lying bound states are Coulombian, but at temperature $T$ they get excited up to
large distances where $V \simeq \sigma r$ ($\sigma \sim \LDC^2$ is the flux tube tension)
forming object with
radius $R_{\B^*} \sim T/\LDC^2$ much larger than the Bohr radius $a_0 = 2/(\adc m_\Q)$.
Writing $V =-\adc/r+ \sigma r$,
a thermal computation gives, for $T<\LDC$
\beq  R_{\B^*}(T)\approx \left( a_0 + \frac{3 m_\Q T^5 \sqrt{m_\Q T}}{\sqrt{\pi} \sigma^4 } e^{-E_B/T} \right)\left( 1+  \frac{ m_\Q T^4 \sqrt{m_\Q T}}{\sqrt{\pi} \sigma^3 }e^{-E_B/T} \right)^{-1}.\eeq
The thermal radius reduces  to $a_0$  for $T\ll E_B$, and to $3 T/\LDC^2$ for $T\sim \LDC$. 
The critical temperature below which the dark baryons relax to the ground state is of order of $E_B$,
and possibly somewhat lower in view of the entropy factor of the almost continuum states of excited states.
At $T\sim \LDC$ an excited baryon $\B^*$ can be approximated as $\ndc$ dark quarks connected by flux tubes with length $R_{\B^*}$.
When $\B^*$ scatters with $\bar \B^*$  two flux tubes can cross:
lattice simulations suggest that the probability of reconnection is close to one;
a similar process takes place in string theory, where the reconnection probability can be suppressed by the string coupling~\cite{hep-th/0412244}.
This results into a large geometric $\sigma_{\B^*\bar \B^*} \sim T^2/\LDC^4$ for $ T  \circa{<}\LDC$,
which enhances $\Q\bar \Q$ annihilations, as their rate inside thermally elongated hadrons is faster than the Hubble rate
(except possibly for hadrons with large angular momenta).
Depending on the precise unknown values of the phase transition temperature $T_c \sim \LDC$ and of 
the string tension $\sigma \sim \LDC^2$ such extra annihilations can be either subleading
or substantially increase the value of the DM mass that reproduces the cosmological DM density~\cite{1606.00159}.
In the rest of the paper we do not consider this possibility.

\section{Signatures}\label{Signatures}

\subsection{Cosmological constraints}

We discuss the various cosmological bounds, that require $\LDC  \circa{>}100$ MeV. 

\subsubsection*{Extra radiation}

If $\Lambda_{\rm DC}\ll 1\MeV~(1\eV)$ dark gluons behave as extra relativistic degrees of freedom at the BBN (CMB) epoch. Their amount can be parametrised 
 as a contribution to the effective number of neutrino species:
\be
\Delta N_{\rm eff}=\frac{8}{7} d(G)\left(\frac{T_{\rm DC}}{T_{\rm SM}}\right)^4
\ee
where $d(G)$ is the dimension of the dark color gauge group.
Present bounds~\cite{1502.01589,1103.1261} constrain $\Delta N_{\rm eff}(T\sim1\MeV)\circa{<}1$ and $\Delta N_{\rm eff}(T\sim1\eV)\circa{<}0.5$. 
This implies 
\be\label{eq:tdgconstr}
 \left(\frac{T_{\rm DC}}{T_\nu}\right)^4=\left(\frac{2}{g_{\rm SM}(T_{\rm dec})}\right)^{4/3}\lesssim\frac{7}{16 d(G)}
\ee
This condition is marginally consistent with $\SU(3)$ and $\SO(3)$ theories if the dark sector decouples at temperature $T_{\rm dec}\gtrsim 1\GeV$.
Models with low confinement scale are however excluded by other cosmological constraints.

\subsubsection*{Structure formation}
Structures such as galaxies form because DM can freely cluster after matter/radiation equality, at $T \circa{<}0.74\eV$.
DM that interacts with lighter dark gluons would instead form a fluid~\cite{1505.03542,1507.04351}:
DM clustering is negligibly affected provided that either
the confinement scale is large enough, 
$\LDC\gtrsim 10\eV$ 
or the dark gauge coupling is small enough,
$\adc\lesssim10^{-8}$.
We will follow the first option.

\subsubsection*{Big Bang Nucleosynthesis}
Dark-glue-balls with mass $\mDG \sim7\LDC$ decay into SM particles
injecting non-thermal particles, which alter the cosmological abundances of light element  or the CMB power spectrum. 
Barring a dark sector with $T_{\rm DC}\ll T_{\rm SM}$, avoiding this requires that injection from glue-ball decays 
is over at the BBN epoch, $T_{\rm SM} \sim \MeV$.
This requires $\LDC \circa{>}\MeV$ and that the dark-glue-ball lifetime $\tauDG$ is shorter than  $1\sec$~\cite{0604251}. 

\subsubsection*{Cosmic Microwave Background}
\begin{figure}[t] 
\begin{center}
\includegraphics[width=0.45\textwidth]{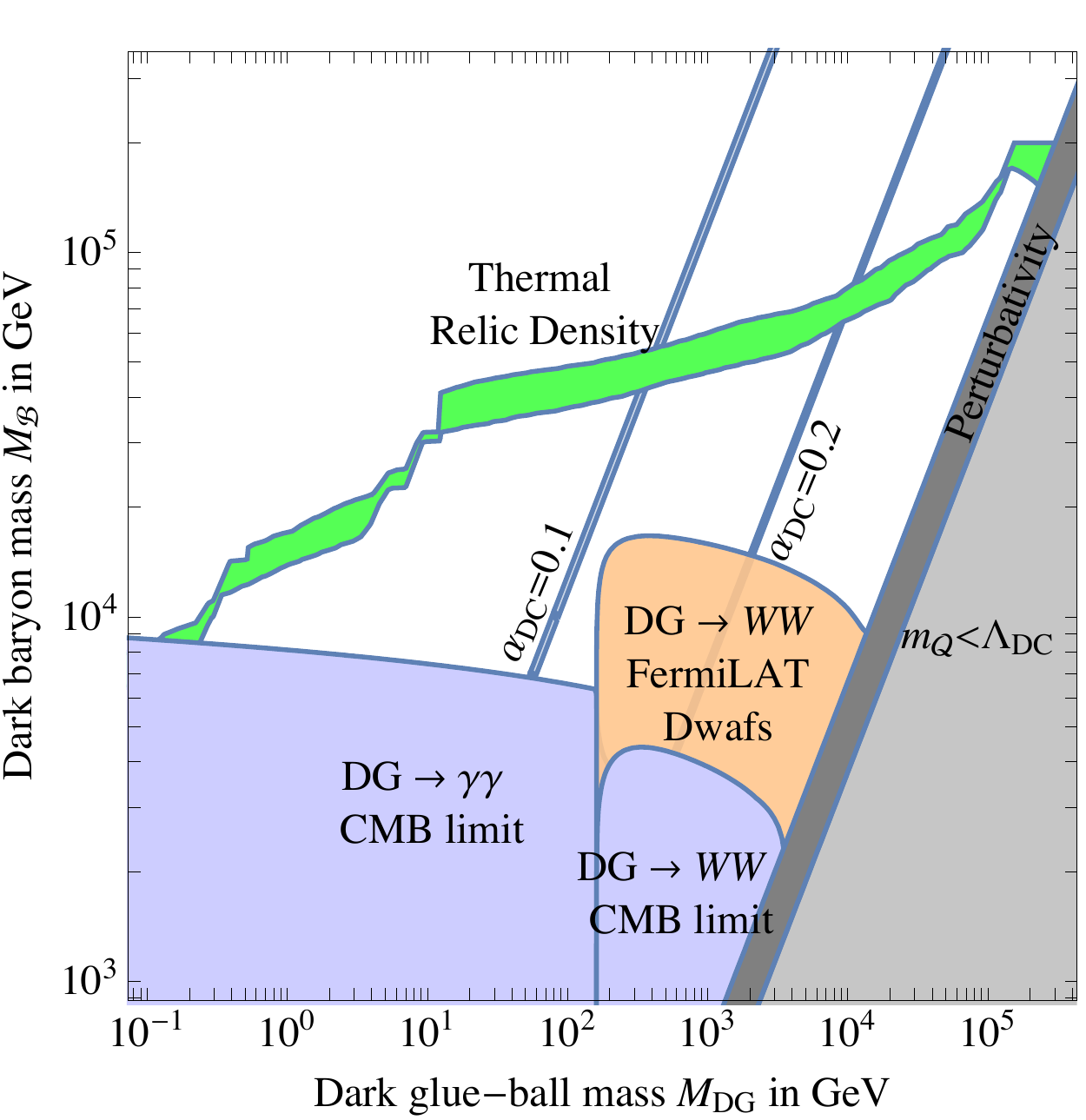}\qquad
\includegraphics[width=0.45\textwidth]{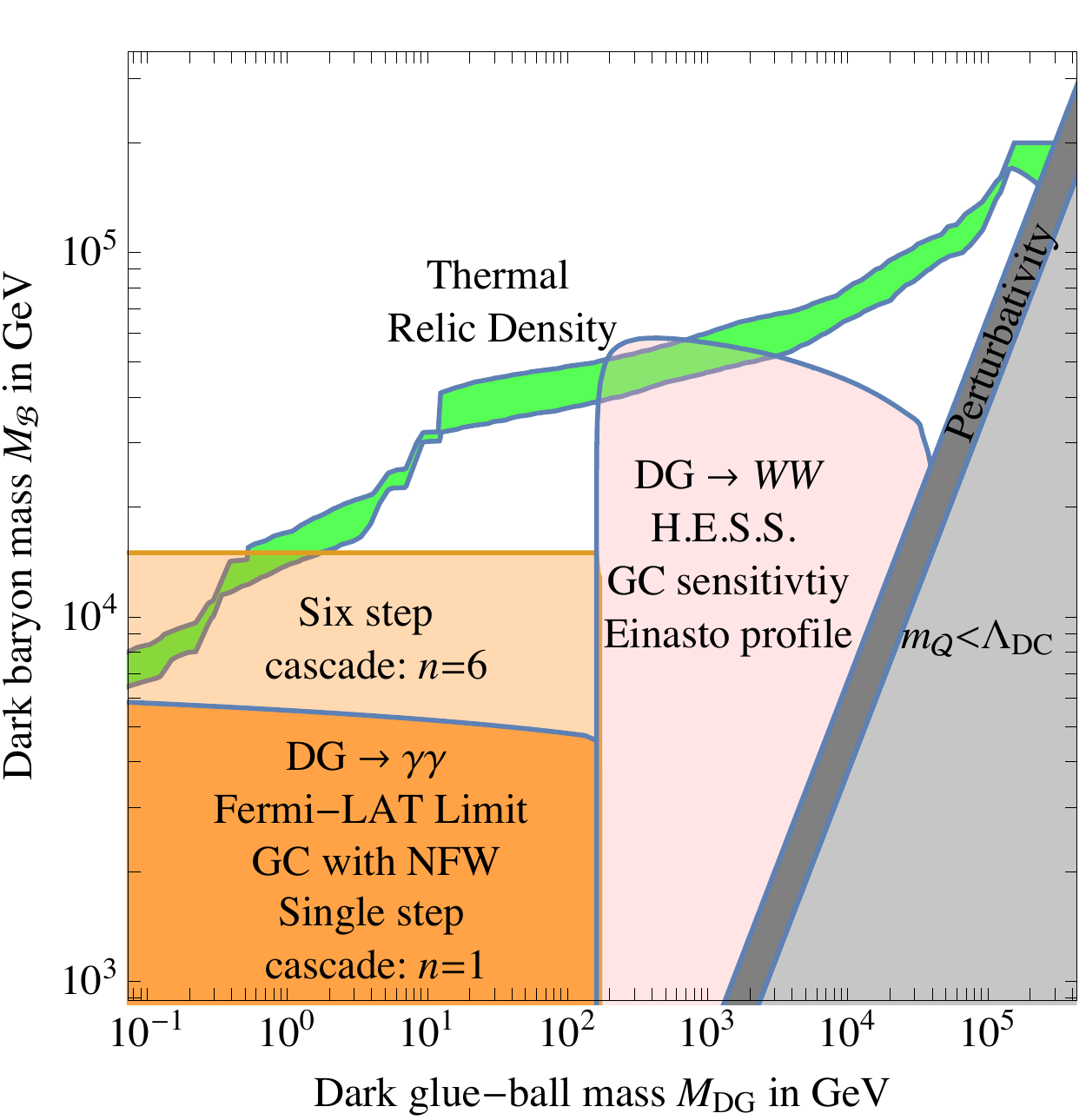}
\end{center}
\caption{\label{fig:CMBglue-balldecay} \label{fig:IndirectV} 
{\bf Left:} \em Indirect detection limits on dark matter annihilation.
Dwarf and CMB bounds (with small astrophysical uncertainties) are compared to cross section values
for dark matter annihilation, which we estimate to be dominated by a recombination reaction of two dark baryons at low velocities.  $M_\B$ is the mass of the DM baryon and $\mDG $ is the  glue-ball mass.
Two regimes are clearly distinguished in the figure, they correspond to either dominant annihilations of glue-balls into $WW$ (for $\mDG  > M_W$) and to dominant annihilations into $\gamma\gamma$  (for $\mDG  <  M_W$).
The green bands show the region where the known DM density is obtained thermally.
{\bf Right}: The sensitivities of Galactic Center observations are considered and the most optimistic DM abundances for indirect detection at the core of the galactic DM profile are assumed. }
\end{figure}

Dark matter that annihilates around photon decoupling at $T_{\rm dec} \sim 0.25\eV$ 
injects particles which ionize hydrogen leaving an imprint on the Cosmic Microwave Background radiation (CMB). 
As the relevant quantity is the total injected power, the CMB bounds on the DM annihilation cross section are robust
and do not depend on the details of the cascade process resulting from DM annihilation to SM final states. 
The bound is weaker than typical indirect detection bounds~\cite{Slatyer}
\begin{align}
\frac{f_\text{eff} \langle \sigma_{\rm ann} v_{\rm rel} \rangle}{\MDM} < 4.1 \times 10^{-28} \frac{\text{cm}^3}{ \text{sec}\,  \text{GeV}} 
\end{align}
where $f_\text{eff}$ is an efficiency parameter depending on the spectra of injected electrons and photons, given by
\begin{align}
f_\text{eff} = \frac{1}{2 \MDM} \int_0^{\MDM} E \,dE \left[  2 f_{\text{eff}}^{e^+ e^-} \left(\frac{dN}{dE}\right)_{e^+} + f_{\text{eff}}^{\gamma} \left(\frac{dN}{dE}\right)_{\gamma}  \right]
\end{align}  
where the ionization efficiencies for $e^\pm$ and $\gamma$ have been computed in~\cite{EfficiencyCMB}. 
In our case $\MDM$ is  the mass of the composite dark baryon. 
The resulting bound is plotted in fig.\fig{CMBglue-balldecay} and leads to a bound on the dark condensation scale $\LDC \gtrsim 30 \text{ MeV} $ in the region where DM is a thermal relic.

\subsection{Indirect detection}\label{indirect}
In the scenario where DM has no dark-asymmetry, dark baryons $\B$ can annihilate 
with dark anti-baryons $\bar \B$ producing indirect detection signals.
The DM kinetic energy $M_\B v^2$ is typically much smaller than the energy of the excited states so that 
we can ignore higher resonances and consider only the ground state dark baryon. Given that after confinement DM is a DC singlet there is no Sommerfeld enhancement due
to DC interactions.  Still, the low-energy annihilation cross section can be large due the large size of the bound states, as discussed in section~\ref{DMBann}, see Eq. (\ref{eq:indirectxsec}).

DM annihilation leads to the production of dark glue-balls, which are the lightest particles in the dark sector. 
The minimal number of produced glue-balls is $k \circa{>}2\ndc$, possibly enhanced up tp
$k\approx\MDM/\mDG $ from dark hadronization effects.
The dark glue-balls later decay to SM particles.  
Dark glue-balls can decay into two photons (if lighter than $M_W$ and of the order of  $\Lambda_{\rm QCD}$)
or --- if $\Q$ is coupled to the higgs ---  into $f\bar f$, where $f$ is heaviest SM fermion lighter than $\mDG/2$.
Details of dark hadronization lead to a characteristically smeared spectrum. 

If the dark glue-ball mass exceeds $2 M_W$ and if dark quarks are charged under $\SU(2)_L$, 
the main decay channel is into two $W$ bosons. 
The decay of the $W$'s leads to a cascade with multiple photons in the final state. 
The electro-weak Sommerfeld corrections are subdominant in comparison to the atomic enhancement of the rearrangement cross sections at low velocities.

In this model framework two possibilities to accomodate for the $e^+$ excess are present. Either the dark glue-balls decay into $\mu^+\mu^-$ and can provide a DM interpretation of
the $e^+$ excess observed by PAMELA and AMS~\cite{PAMELA} for TeV scale dark baryons or dark glue-balls decay to $W^+ W^-$ and explain the excess if the mass of the baryons if $M_\B > 10 \text{TeV}$. The annihilation cross section is large thanks to the $\B\bar\B$ cross section enhancement by recombination.

\subsection{Direct Detection}
\label{sec:dd}

Direct detection experiments see DM dark-baryons as a particle and cannot resolve its constituents.
Indeed, the maximal momentum transfer in elastic interactions with nuclei of mass $m_N$
is  $\approx m_N v \circa{<}100\MeV$ in view of the galactic DM velocity $v\sim 10^{-3}$.
In the range of parameters allowed for our models the size of DM bound states is smaller than the corresponding wave-length 
so DM bound states scatter coherently with the nucleus.\footnote{Some fraction of dark baryons could form dark nuclei~\cite{Nuclei}, affecting direct detection signals.}. 

\subsubsection*{$\SU(\ndc)$ models}
We first discuss $\SU(\ndc)$ models where DM is complex.
In the simplest case the dark-baryon DM belongs to a single multiplet of the SM interacting as in minimal dark matter models~\cite{hep-ph/0512090}. 
Direct detection constraints on $Z$-mediated scatterings are satisfied if the DM candidate has no hyper-charge, which implies integer isospin. 
The loop-level $W$-mediated cross section~\cite{hep-ph/0512090,0706.4071,Hill:2013hoa}
is independent of the dark matter mass and entirely dependent by its $\SU(2)_L$ quantum number,
equal to about $\sigma_{\rm SI}\approx 1.0 \times 10^{-45}\cm^2$ for a weak triplet,
and to $\approx 9.4 \times 10^{-45}\cm^2$ for a weak quintuplet.
The predicted cross-sections are above the neutrino floor and will be observable in future experiments if $M_\B \lesssim 15 \TeV$.

\medskip

This simple result can however be drastically modified in the presence of heavier dark fermions. 
In models where the DM fermion has Yukawa couplings ($y$ for the left-handed chirality and $\tilde{y}$ for the right-handed chirality)
with the Higgs and with an heavier dark-quark with non vanishing hypercharge, 
the DM candidate can acquire  a vector coupling to the $Z$.  The heavier dark-quarks  have a vectorial coupling to the $Z$ given by
\begin{equation}
g_Z\equiv \frac {g_2} {\cos\theta_{\rm W}}\left(T_3-Q\sin^2\theta_{\rm W}\right).
\end{equation}
After electro-weak symmetry, the dark-quarks that make up the DM mix with the heavier dark quarks, acquiring 
an effective vectorial coupling
\begin{equation}
g_Z^{\rm eff}=\frac{g_Z}2 (s_L^2+s_R^2)
\end{equation}
where $s_L$ and $s_R$ are the mixing of left and right chiralities. Since the $Z$ is coupled to a conserved current, 
the coupling $g_Z^\B$ to dark baryons is given by the sum of the constituent charges. 
For example $g_Z^\B= N_{\rm DC} g_Z^{\rm eff}$ when the dark-baryon is made of electroweak singlets. 
At low energies we obtain the effective interaction between $\B$, the DM dark baryon, and the SM quarks $q$
\be
\La_{\rm eff}\supset  \frac {g_Z^\B g_Z^q}{M_Z^2}(\bar \B\gamma^\mu \B)( \bar q\gamma_\mu q).
\ee
From this Lagrangian one obtains the spin-independent DM cross section on nuclei $N$
\be
\sigma_{\rm SI}=\frac{(\mu_n G_{\rm F}\cos\theta_{\rm W})^2}{4 \pi}\left(\frac{g_Z^\B}{g_2}\right)^2
\ee
where  $\mu_n$ is the reduced mass of the DM-nucleon system.
The direct detection bound implies $g_Z^\B\lesssim 7\times10^{-4}\sqrt{M_\B/\TeV}$.

\medskip

When Yukawa couplings exist, Higgs mediated scatterings are also generated. 
The Yukawa coupling to the lightest mass eigenstate is
$
y_{\rm eff}= y s_L c_R +\tilde{y} c_L s_R
$.
The Yukawa coupling of dark-baryons is given by the sum of the Yukawa of the constituent dark quarks.
The resulting SI cross section is~\cite{DelNobile:2013sia}:
\be
\sigma_{\rm SI}=\frac {\sqrt{2} G_F   f_n^2}{\pi } \frac{\mu_n^4}{M_h^4}y_\B^2
\ee
where $f_n\approx 1/3$ is the relevant nuclear form factor \cite{1110.3797,1209.2870}. Direct detection bounds imply $y_\B\lesssim4\times10^{-2}\sqrt{M_\B/\TeV}$.

Furthermore, fermionic composite DM that contains electrically charged constituents has
a magnetic moment $\mu \sim e \adc/(4\pi) m_\Q$ that can lead to a potentially observable cross-section
with characteristic dependence on the recoil energy $E_R$, $d\sigma/dE_E \approx e^2 Z^2 \mu^2/4\pi E_R$.

\subsubsection*{$\SO(\ndc)$ models}
Models based with dark quarks in the fundamental of  $G_{\rm DC}=\SOD$   behave differently, because 
the lightest fermion is a real Majorana state that cannot have  vectorial couplings to the $Z$.
Mass eigenstates $\chi_M$  have only axial couplings to the $Z$ 
\begin{equation}
\tilde{g}_Z^{\rm eff}\,\bar{\chi}_M \gamma_\mu\gamma_5 \chi_M\qquad\hbox{with strength}\qquad
\tilde{g}_Z^{\rm eff} =\frac{g_2}{2\cos\theta_{\rm W}} (s_L^2-s_R^2).
\end{equation} 
This contributes to spin dependent cross-sections with nuclei, subject to much weaker bounds.
For this reason DM candidates with non-zero hypercharge are possible in the presence of a small mixing with a real particle.
For what concerns Higgs interactions these are as in $\SU(\ndc)$ and similar bounds apply.

Vector coupling to the $Z$ can be present between DM and heavier states.
 DM made of 
electro-weak doublets gives two almost degenerate Majorana fermions split by
\begin{equation}
\Delta m \sim \frac {y^2 v^2}{\Delta \mQ}
\end{equation}
where $\Delta\mQ$ is the mass splitting between the two dark quarks which get mixed. 
When the splitting is smaller than $\mathcal{O}(100\KeV)$ inelastic transitions between the two states 
can take place giving rise to inelastic dark matter~\cite{TuckerSmith:2001hy}. 

\medskip

Finally, we comment on dipole moments.
In models with $G_{\rm DC}=\SU(\ndc)$ and $\mQ \ll \LDC$,
fermionic baryons acquire large magnetic dipole moments
(which give characteristic signals
in direct detection experiments~\cite{Antipin:2015xia}) 
thanks to non perturbative effects.
If instead $\mQ\gg\LDC$, neutral baryons have small magnetic moments given (at leading order) by the sum of the
elementary moments. A similar result holds for electric dipoles, possibly generated by a $\theta_{\rm DC}$ angle
by instantons, which are suppressed in the perturbative regime. {Polarisability of weakly coupled dark matter bound states 
could also be of interest \cite{1609.01762,1503.04205}.}

\subsection{Collider}
If dark quarks are charged under the SM, bound states of the new sector can be  produced singly or through 
the hadronization of the dark quarks  produced in Drell-Yan processes. 

Resonant single production does not depend on the details of the strong dynamics.
In the narrow width approximation, the production cross-sections of a  bound state $X$ of mass $M_X$ is given by
\begin{equation}
 \sigma(pp\to X) =\frac{(2 J_X+1) D_X}{M_X s} 
\sum_{\mathcal P} C_{{\mathcal{P P}}}  \Gamma(X\to \mathcal {P P}) \,,
\label{eq:singleproduction}
\end{equation}
where $D_X$ is the dimension of the representation, $J_X$ is its spin, $\mathcal P$ the parton producing the resonance
and $C_{{\mathcal{P P}}}$ are the dimension-less parton luminosities, see~\cite{1512.04933}. 

\begin{figure}[t]
\begin{center}
\includegraphics[width=0.45\textwidth]{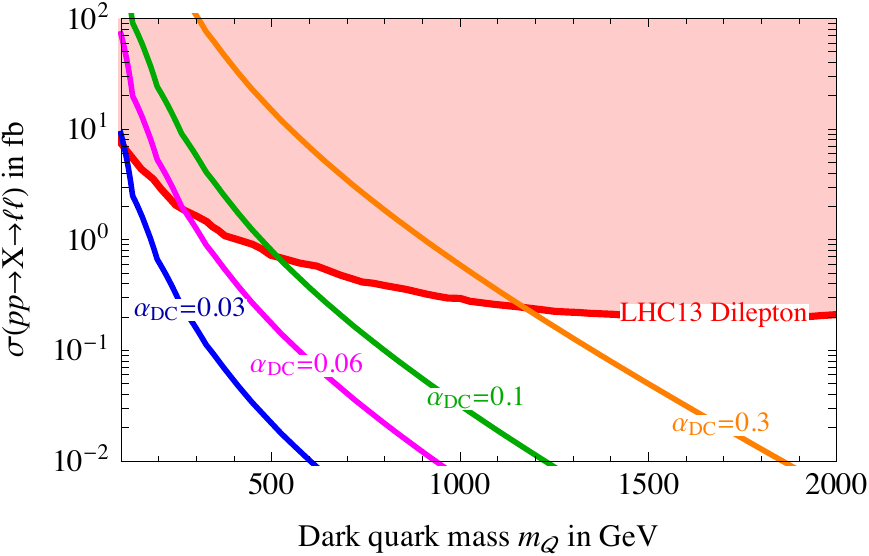}\qquad
\includegraphics[width=0.45\textwidth]{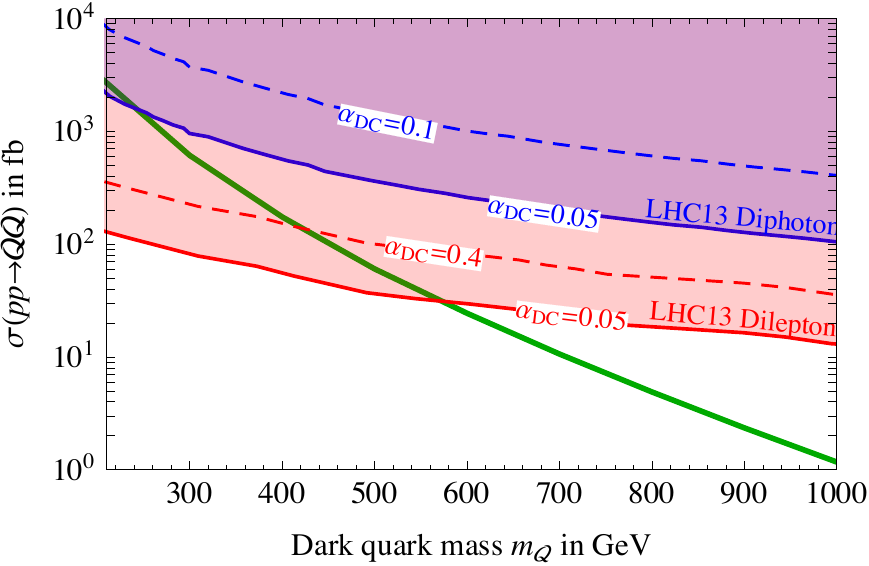}
\caption{\label{fig:collbound}\em {\bf Left:} ATLAS bounds on the cross section for the direct production of a spin 1 resonance decaying into leptons ($\mu$ und e)~\cite{ATLAS:2017wce}. {\bf Right:} ATLAS bounds on the dark quarks pair production cross section ~\cite{ATLAS:2017wce}. They are derived assuming that $\sim1/3$ of the produced dark quarks form spin $1$ bound states and the others spin $0$ bound states.}.
\end{center}
\end{figure}

Bound states with spin-0 are produced from vector bosons fusion. For constituent dark quarks  
with $\SU(2)_L \times \U(1)_Y$ quantum numbers the decay width of singlet spin-0 bound states is
\be\label{eq:rates0gg}
\Gamma\left(X^{J=0}_{I=0}\to\gamma\gamma\right)=N_{\rm DC}\alpha^2\frac {|R_{n0}(0)|^2}{F^2 \mQ^2} \frac {(T_2+d_2 Y^2)^2}{d_2} 
\ee
where $T_2$ ($d_2)$ is the index (dimension) of the $\SU(2)_L$ representation, $R_{n0}(0)$ is the 
value at the origin of the bound state wave-function and $F=1(2)$ for distinguishable (identical) dark quarks.  The decay rates into $W$ and $Z$ bosons and
into dark gluons $\cal G$ are 
\begin{equation}\label{eq:rates0}
\begin{array}{rclrcl}\displaystyle
\frac {\Gamma_{\gamma Z}}{\Gamma_{\gamma\gamma}}&\approx& \displaystyle
\frac {2(- T_2 \cot \theta_{\rm W}+ Y \tan \theta_{\rm W})^2}{(T_2+d_2 Y^2)^2}\,,\qquad &\displaystyle
\frac {\Gamma_{Z Z}}{\Gamma_{\gamma\gamma}}&\approx&\displaystyle
\frac {(T_2 \cot \theta^2_W+ Y \tan \theta^2_W)^2}{(T_2+d_2 Y^2)^2},\\\\  \displaystyle
\frac {\Gamma_{WW}}{\Gamma_{\gamma\gamma}}&\approx&  
\displaystyle
2 \frac{T_2^2}{(T_2+d_2 Y^2)^2 \sin^4 \theta_{\rm W}}\,, \qquad &\displaystyle
\frac {\Gamma_{{\cal GG}}}{\Gamma_{\gamma\gamma}}&\approx&  
\displaystyle
\frac 1 {16 F} \frac{N_{\rm DC}^2-1}{N_{\rm DC}^2}\frac {d_2^2}{(T_2+d_2 Y^2)^2}\frac{\adc^2}{\alpha^2}\,.
\end{array}
\end{equation}
Spin-1 bound states decay into fermions or scalars (and equivalent longitudinal gauge bosons $W,Z$),
as their decays into massless gauge bosons is forbidden by the Landau-Yang theorem.
For example, the decay width of  an $\SU(2)$ triplet spin-1 bound state into a left-handed pair of SM fermions is 
\be\label{eq:rates1ff}
\Gamma^a \left(X^{J=1}_{I=1}\to f\bar f\right)=N_{\rm DC} \frac{\alpha^2_2 }{12}\frac{\left|R_{n0}(0)\right|^2}{F^2 \mQ^2}T_2
\ee
where we neglected possible hypercharge contributions. Singlet spin-1 bound states can also decay into three dark gluons with a rate:
\be\label{eq:rates1GGG}
\Gamma_{{\cal GGG}}=N_F \frac{\sum_{abc}d_{abc}^2}{36 \, d_R}\frac{\pi^2-9}{\pi}\adc^3\frac{\left|R_{n0}(0)\right|^2}{F^2 \mQ^2}
\ee
where $d_{abc}=2 \Tr\left[T^a\{T^b,T^c\}\right]$ with $T^{a,b,c}$ generators of the dark-color group in the dark quarks representation.

\medskip

For concreteness we focus on the model with $G_{\rm DC}=\SU(3)$  with a dark quark $\Q=V$.  In the region of parameters relevant for DM,
the dark coupling $\adc$ is stronger than the electro-weak couplings, so that the bound states are dominantly shaped by the dark interactions. 
In the Coulomb limit, the radial wave function at the origin is then given by $|R_{n0}(0)|^2/\mQ^2=(F \mQ\alpha_{\rm eff}^3)/(2n^3)$ with $\alpha_{\rm eff}$ defined as in \eqref{eq:coulombpot}. Spin-0 bound states are produced from photon fusion and decay mostly into dark gluons with the branching ratios given
in eq.~(\ref{eq:rates0}). In view of the small photon luminosity at LHC, no significant bound is obtained. 
Spin-1 resonances can be produced in electro-weak interactions from  first generation quarks and decay into electrons and muons with a branching ratio
of order $15\%$, neglecting decays to 3 dark gluons. In fig.~\ref{fig:collbound} we show the bound from current di-lepton searches that exclude dark quark masses up to 1 TeV.
This is significantly stronger than typical collider bounds on electro-weak charged states.

\medskip

Dark quarks with SM charges can be also pair produced in Drell-Yan processes. 
In the region of masses relevant for LHC, their kinetic energy is comparable to their mass. 
When dark quarks travel a distance $\ell \gg 1/\LDC$  a flux tube develops between them carrying an energy $\LDC^2 \ell$,
such that they reach a maximal distance~\cite{0805.4642} 
\begin{equation}
\ell_{\rm max}\sim \frac {\mQ}{\LDC^2}\sim 10^{-13} \,{\rm m} \bigg(\frac {\mQ}{\rm TeV}\bigg)\bigg(\frac {\rm GeV}{\LDC}\bigg)^2
\end{equation}
which is microscopic in the region relevant for DM phenomenology.
The dark quarks will then oscillate and de-excite to the lowest lying bound states with the emission of dark glue-balls, until they eventually decay to SM states. 
It is difficult to determine the branching ratios into each SM channel. 
Assuming for simplicity that all dark quark pairs de-excite democratically to the lowest lying spin-0 and a spin-1
bound states, 2/3 of the events populate the spin-0 bound states (singlet and quintuplet) and 1/3 populate
the spin-1 triplet. In fig.~\ref{fig:collbound} we show the bounds from di-photons and di-leptons on
double productions of dark quarks. Especially in the region of large $\adc$, these bounds are weaker than the bounds from single production.

\medskip

\subsection{Dark glue-balls at high-intensity experiments}\label{DG2}
Dark glue-balls can be produced either through the production and subsequent decay of dark mesons 
or through the effective operators~\cite{Juknevich:2009ji,0911.5616,Forestell:2016qhc}
\begin{equation}
{\cal O}_8= \alpha_{\rm em} \adc {\cal G}_{\mu\nu}^A{\cal G}^{\mu\nu A}{F}^{\rho\sigma}{F}_{\rho\sigma}\,, 
\quad \quad {\cal O}_6=
\frac{\adc }{4\pi}H^\dagger H {\cal G}_{\mu\nu}^A{\cal G}^{\mu\nu A}
\label{eq:dim8}
\end{equation}
The diagrams in fig.~\ref{fig:tcballdecay} generate ${\cal O}_{6,8}$ with  coefficients
\begin{equation}
c_8(\mQ)=\frac {T_{\rm DC}(T_2+d_2 Y)}{60}\frac 1 {\mQ^4}\,, \qquad c_6(\mQ)= \frac {2 T_{\rm DC}} 3\frac 1  h \frac{\partial \ln (\det M_F(h) )]}{\partial h}\bigg|_{h=0}
\end{equation}
where $M_F(h)$ is Higgs-dependent dark quark mass matrix, 
$T_{\rm DC}$ the index of the dark quark, 
$T_2$ the isospin, and $Y$ its hypercharge.

After confinement, ${\cal O}_{8}$ gives rise to a coupling between $0^{++}$ glue-balls and the SM gauge bosons which allows  the glue-balls  
to decay into photons. For the lightest $0^{++}$ glue-ball one finds~\cite{0911.5616}
\begin{equation}
\Gamma_{0^{++}\to \gamma \gamma}= \frac {\alpha_{\rm em}^2 \adc^2}{14400 \pi} \frac {m_0^3 f_{0S}^2}{\mQ^8} (T_2+d_2 Y^2)^2
\end{equation}
where $f_{0S}\equiv \langle 0 | {\rm Tr}\,{\cal G}_{\mu\nu} {\cal G}^{\mu\nu}| 0^{++}\rangle$. 
Using the lattice result $4\pi \adc f_{0S}\approx 3 \mDG ^3$ valid for $\SU(3)$ theories, one gets the dark-glue-ball lifetime in eq.~(\ref{eq:Glifetime}) for models with electro-weak charges. 
The Yukawa couplings between the dark and the SM sector  induce a mixing angle $\alpha$ between dark glue-balls and the SM Higgs
\begin{equation}
\sin \alpha \approx c_6 \frac {\adc}{4\pi} \frac{v f_{0S}}{M_h^2}
\end{equation}
giving rise to the dark glue-ball decay widths 
\begin{equation}
 \Gamma_{0^{++}\to f\bar f}= N_c\frac {\mDG}{16 \pi} y_f^2 \sin^2 \alpha \,, \qquad  \, \Gamma_{0^{++}\to gg}= \frac{\alpha_s^2}{72\pi^3}\frac{\mDG^3}{v^2} \sin^2\alpha \,, 
\end{equation}

\medskip

\begin{figure}[t!]
\begin{center}
\includegraphics[width=.6\textwidth]{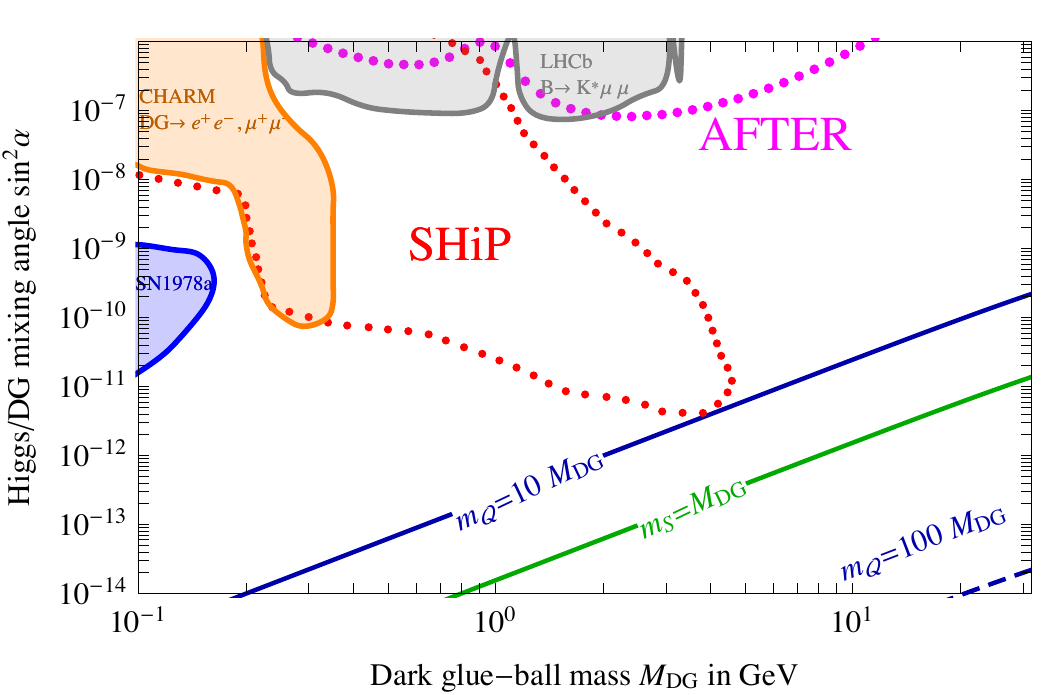}
\caption{\em  Predicted values of the
Higgs/dark gluon mixing angle $\alpha$, assuming dark quarks with
Yukawa couplings $y=1$ (blue lines) or adding a dark  scalar with mass $m_S$ (green line) with a mixed quartic $\lambda_{HS}=10^{-2}$, as function of the dark gluon mass $\mDG$.
The shaded regions are excluded, the dotted curves can be probed by future SHiP~\cite{1504.04855} (red points) and AFTER  \cite{1612.06769} (magenta points) experimental proposals.
\label{fig:ship}}
\end{center}
\end{figure}

The cross-section for the production of dark glue-balls are negligible at LHC.
Light dark glue-balls can be potentially produced in high luminosity experiments such as SHiP~\cite{1504.04855}.
The SHiP experiment will operate at a center of mass energy $E_{\rm CM}\approx 27\GeV$ and will produce approximately $10^{20}$ proton on target  collisions.
The distance from the target to the detector is approximately $L\sim 100\,{\rm m}$ and the detector length is $S\sim 60\,{\rm m}$. 
A detectable signal at SHiP arises if there are a few events in the detector 
\be N_{\rm ev}\sim
10^{20}\frac{\sigma(pp\to {\rm DG})}{\sigma_{pp}}\times\left[e^{-L/\tau_{\rm DG}}\left(1-e^{S/\tau_{\rm DG}}\right)\right]\circa{>} \hbox{few}
\ee
where   $\sigma_{pp}\sim1/m_p^2$ is the proton-proton scattering cross section. 
This implies that the SHiP experiment will probe only a region of the parameter space which is already excluded by indirect detection bounds or electroweak precision tests (see next section). This conclusion is confirmed by the result of a more precise computation, shown in fig.\fig{ship}. In the same figure we also show the sensitivity of an hypothetical fixed target experiment (AFTER) operating 
with LHC beams at a center of mass energy $E_{\rm CM}\approx 115\GeV$ and producing approximately $10^{15}$ proton on target~\cite{1612.06769}. 

The conclusion persists even if the theory is modified by adding an extra dark colored scalar neutral under $G_{\rm SM}$,
coupled to the Higgs as $\lambda_{HS} |S|^2 |H|^2$,
which gives an extra contribution  $c_6= \lambda_{HS} T_{\rm DC}/(12 m_S^2)$, enhanced by its possibly small mass
$m_S\ll M_h$. Imposing $|\lambda_{HS}|\circa{<}0.01$ in view of bounds on the Higgs invisible width,
and $m_S \circa{>}\LDC$ in order not to change the DM phenomenology, we find that dark glue-balls remain undetectable at SHiP.

\subsection{Radioactive Dark Matter}\label{sec:rad}
\begin{figure}[t] 
\begin{center}
\includegraphics[width=0.45\textwidth]{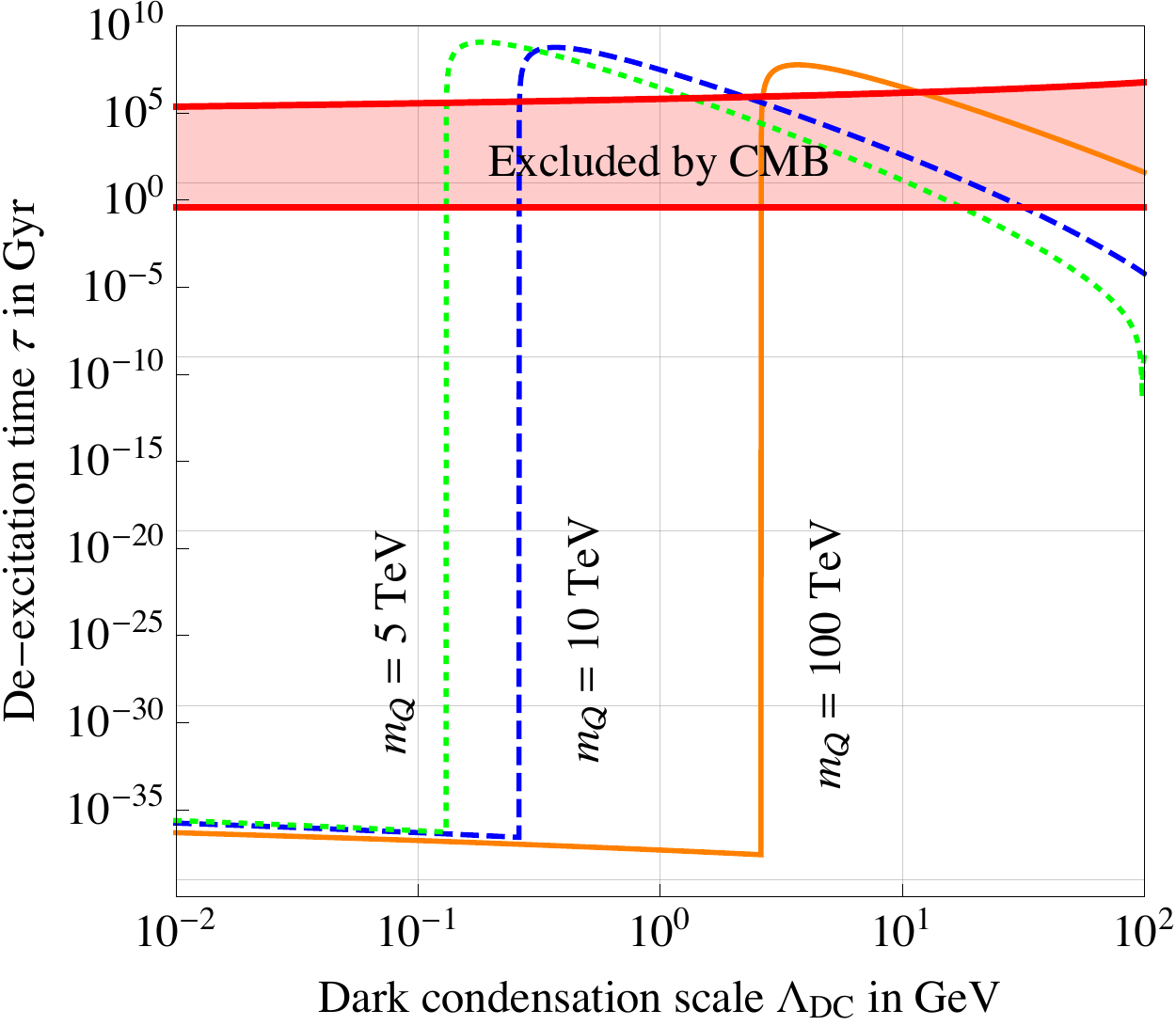} \quad
\includegraphics[width=0.45\textwidth]{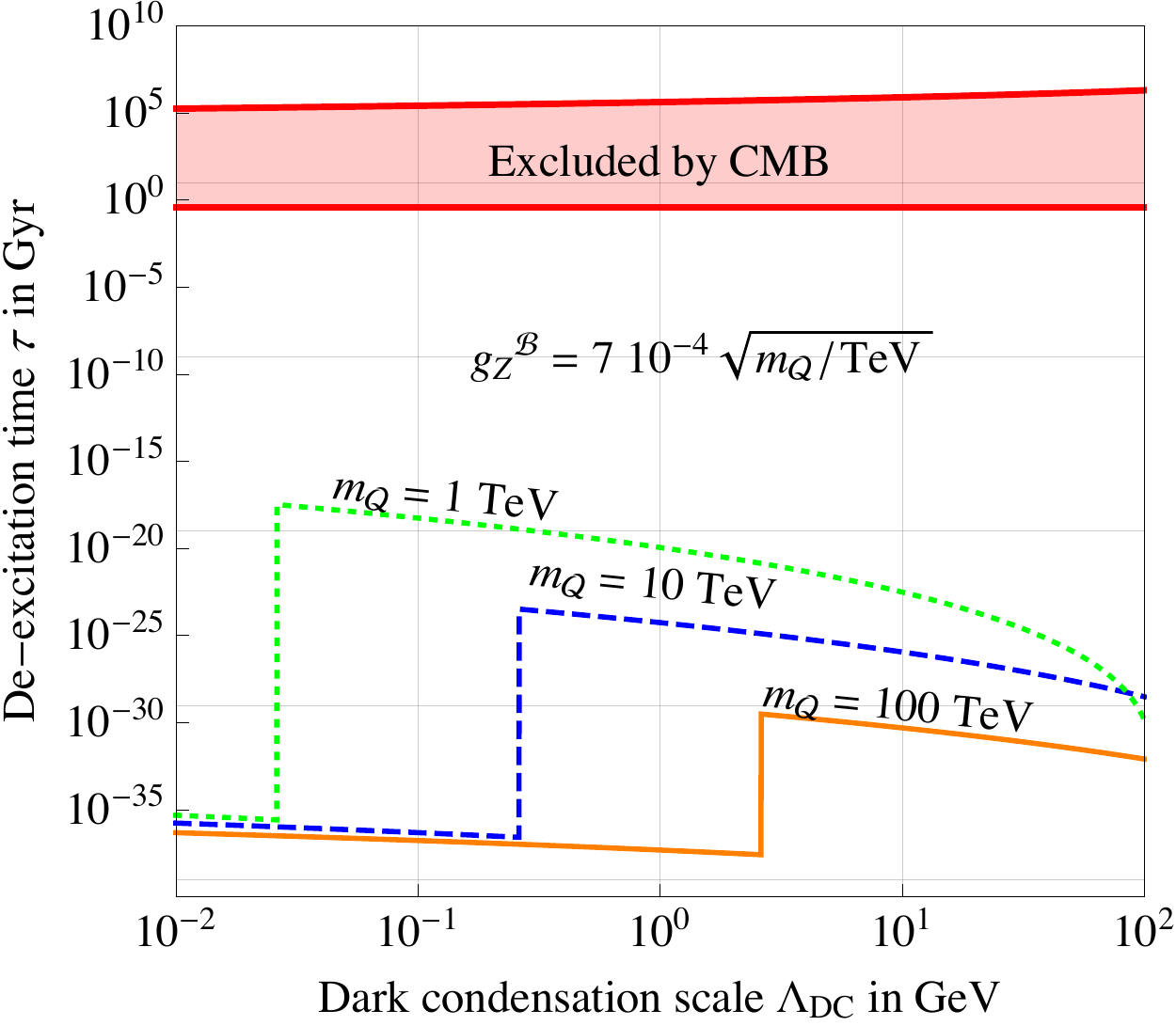}
\end{center}
\caption{ \label{fig:decaytime} 
\em De-excitation life-time of radio-active dark matter,
that can be long when $\mDG \sim 7 \LDC$ is larger than the binding energy.
A very long $\tau$ is obtained when the life-time of dark glue-balls is so long that they dilute the DM density.
In the left (right) panel glue-balls decay thanks to heavier dark quarks charged under $G_{\rm SM}$
(with a Yukawa coupling $y$ to the Higgs).
Bounds from energy injection in the CMB spectrum are shown.}
\end{figure}
As discussed in section~\ref{bound} the parameter space allows for $E_B   \circa{<} \LDC  \circa{<} 1/a_0$ (region B).
This leads, in the primordial universe at temperatures $T \circa{<} \LDC$, to the production of excited DM bound states.
These states  can be long-lived if $\Delta E_B <\mDG$ such that decays to a dark glue-ball are kinematically forbidden.
In models where $\Q$ is neutral under the SM, such excited bound states then can only decay to light SM states 
(such as $\gamma\gamma$ or $e^+e^-$)  through an off-shell glue-ball-like state,  giving rise to {\em radioactive dark-matter}.
We can estimate the decay rate of such trapped excited bound states, by splitting the phase space in terms of the invariant mass $M$ of
the off-shell virtual dark glue-ball DG \cite{1512.04933}, obtaining
\begin{eqnarray} 
\Gamma(\B^* \rightarrow \B\,{\rm SM})& = & \frac{1}{\pi}  \int_0^{\Delta E_B^2} M d M^2\,  \frac{\Gamma (\B^* \rightarrow \B\, {\rm DG}^*(M) ) \, \Gamma_{\text{DG}^*}(M) }{|M^2-\mDG^2 + i \GammaDG \mDG|^2}
\label{eq:integrated}\,.
\end{eqnarray}
where $\Gamma (\B^* \rightarrow \B\,{\rm DG}^*) $ is the decay width into a virtual dark glue-ball with mass $M$,
and $\Gamma_{\text{DG}^*}(M) $ is its decay width into SM states. We approximated the imaginary part of the propagator $M \Gamma_{\text{DG}^*}(M)$
with the value on-shell. If the dark glue-ball can be on shell, the integral around its peak gives 
$\Gamma(\B^* \rightarrow \B\,{\rm SM})\simeq \Gamma (\B^* \rightarrow \B\, {\rm DG}^* )$.
We are interested in the opposite regime where the intermediate state is off-shell. For  $\mDG \gg \Delta E_B$ the propagator is approximately 
constant and we estimate
\beq\Gamma(\B^* \rightarrow \B\,{\rm SM})\sim  \frac{\Delta E_B^3}{\pi \mDG^4}    \Gamma (\B^* \rightarrow  \B\, {\rm DG}^*(0))\Gamma_{\text{DG}^*}(\Delta E_B) \,.\label{eq:abovefor}
\eeq
Taking into account that ${\rm DG}^*$ is a dark glue-ball-like state that does not need to have spin 0, but
can match the quantum numbers of two dark gluons, we estimate $\Gamma (\B^* \rightarrow  \B\, {\rm DG}^*(0)) \approx \adc^4( m_\Q)\adc^2(\LDC) m_\Q$ 
as the decay rate into two massless dark gluons. The 4 powers of $\adc( m_\Q)$ arise from the bound-state wave function and binding energy, 
while the two powers of $\adc(\LDC)\sim 1$ arise from dark-gluon emission.
$\Gamma_{{\rm DG}^*}(\Delta E_B)$ can be small, making  excited $\B^*$ long lived, as shown in fig.\fig{decaytime},
where the large increase of the life-time corresponds to the transition from on-shell to off-shell decays. In models where Yukawa couplings exist excited DM can decay through $Z-$mediated processes giving a much shorter lifetime, see fig \ref{fig:decaytime} right panel.

Bounds on radioactive DM can be inferred by rescaling bounds on decaying DM.
An excessive reionization of CMB is roughly
obtained for $t_{\rm CMB} \ll \tau < 1.1\, 10^9 \,{\rm Gyr}\times  \Delta E_B/M_\B$~\cite{1504.01195},
where $t_{\rm CMB} \approx 380\,{\rm kyr}$ is the Universe age at photon decoupling and $M_\B$ is the DM mass.
If DM is still $\gamma$-radioactive today, one must have
$\tau > 10^{11}\,{\rm Gyr}\times  \Delta E_B/M_\B$, for $0.1\MeV \circa{<}\Delta E_B \circa{<}10\GeV$~\cite{radioactiveDM}.
If DM is still $\beta$-radioactive today, its de-excitation life-time(s) must be longer than
$\tau > 10^{7}\,{\rm Gyr}\times  \Delta E_B/M_\B$, for $\MeV \circa{<}\Delta E_B \circa{<}10\GeV$~\cite{radioactiveDM}.
DM with $\tau \sim T_U$ can be borderline at MeV.
In view of these bounds and of the model predictions,
 its seems unlikely that DM can be radioactive enough to heat
solving the small-scale potential `cusp/core' and `missing satellite' problems of cold DM.
In the parameter region without dark matter dilution by glue-ball decay the glue-ball lifetime has to be smaller than one second, as we  discussed earlier. This leads to a half life of the radiative states of the order of a few hours. Thus they have no impact on the CMB spectrum.

\subsection{Precision tests}
Vector-like fermions do not give large corrections to electro-weak precision observables. 
The regime $\mQ\ll\LDC$ was discussed in~\cite{1508.01112,1609.07122}.
The result in the opposite regime is qualitatively similar. 
The corrections to the precision $S$ and $T$ parameters are
\begin{equation}
\Delta \hat{T}  \sim N_{\rm DC} \frac{y^4}{16 \pi^2} \frac{v^2}{\mQ^2},\qquad
\Delta \hat{S} \sim N_{\rm DC}\frac{y^2}{16\pi^2}\frac{v^2}{\mQ^2}\,.
\end{equation}
Experimental bounds allow  couplings $y\sim 1$ if $\mQ$ is above a few hundred GeV.

\medskip

Extra Yukawa coupling can give extra effects in flavour. 
For general Yukawa couplings, the 
theory contains CP violating phases
$
{\rm Im}[m_{\Q_1} m_{\Q_2} y^* \tilde{y}^*]$
which generate electric dipole moments of SM particles at two loops. 
Similar effects have been studied in supersymmetry~\cite{0510197}.
In a model with $\Q = L\oplus V$ we estimate
\begin{equation}
d_f \sim  N_{\rm DC} e Q_f \frac {\alpha\, {\rm Im}\, y \tilde{y}} {16 \pi^3} \frac  {m_f}{m_L m_V} \ln\frac {m_L m_V}{M_h^2} \,.
\end{equation}
For the electron this means
\begin{equation}
 {d_e}\sim  10^{-27}\, e\, {\rm cm}\times    {\rm Im}[y \tilde{y}] \times \frac{N_{\rm DC}}{ 3} \times\frac {\rm TeV^2}{m_L m_V}
\end{equation}
to be compared with the experimental bound $d_e<8.7 \times 10^{-29}~e\,{\rm cm}$~\cite{1310.7534}. 
A somewhat smaller effect is obtained in the $\Q=L\oplus N$ model.

\section{Models}\label{Models}
Finally, we analyse the microscopic structure of the simplest models with $\SU(\ndc)$ and $\SO(\ndc)$ dark gauge interactions.
At energies greater than $\LDC$ we have a set a fermions charged under $G_{\rm DC}\otimes G_{\rm SM}$. They annihilate into SM
degrees of freedom or dark gluons. Moreover they can form bound states through the emission of dark gluons or SM gauge bosons.

At tree level, a dark quark with mass $m_\Q$ has the following
$s$-wave annihilation cross section into massless gauge bosons
\begin{equation}
\langle\sigma v_{\rm rel} \rangle_{\rm ann}=\frac {A_1+A_2}{16\pi g_{\chi} d_R}\frac{1}{\MDM^2}
\end{equation}
where
\begin{equation}
A_1\equiv {\rm Tr}[T^a T^a T^b T^b]\,,\qquad
 A_2\equiv {\rm Tr}[T^a T^b T^a T^b]
\end{equation}
and $g_\chi=4(2) d_R$ for Dirac or Majorana fermions. For dark quarks charged under both $G_{\rm DC}$ and $G_{\rm  SM}$ the  notation above 
stands for $T\equiv (g_{\rm DC} T_{\rm DC} \otimes 1) \oplus  (1 \otimes g_{\rm SM} T_{\rm SM})$.
For dark quarks in the irreducible representation $(N,R_{\rm SM})$ the formula above gives
\begin{equation}
\langle\sigma v_{\rm rel} \rangle_{\rm ann}=\frac 1 {d_{\rm SM}}\frac {K_1^{\rm DC}+K_2^{\rm DC}}{4(2) N_{\rm DC}^2 }\frac{\pi \alpha_{\rm DC}^2}{\MDM^2} +\frac 1 {N_{\rm DC}}\frac {K_1^{\rm SM}+K_2^{\rm SM}}{4(2) d_{\rm SM}^2 } \frac{\pi \alpha_{\rm SM}^2}{\MDM^2}+\frac {4 C_{\rm DC} C_{\rm SM}}{4(2) d_{\rm SM} N_{\rm DC} }\frac{\pi \alpha_{\rm DC}\alpha_{\rm SM}}{\MDM^2}\,.
\label{eq:annDCSM}
\end{equation}
The group theory factors are listed in table~\ref{tab:Ccolor} using
\begin{equation}
K_1(R)= d(R) C(R)^2\,,\qquad
K_2(R)=K_1(R)-\frac {d(A)C(A)T(R)}2\,.
\end{equation}
Furthermore, 
dark quarks charged under the SM undergo extra annihilations into SM fermions and into the Higgs.

\begin{table}[t]
\begin{center}
\begin{tabular}{c|c|c}
\rowcolor[HTML]{C0C0C0} 
$G_{\rm DC}$ & $R_i\rightleftarrows R_f$ & $I_{R_i\to R_f}$\\ \toprule
                           & $1\rightleftarrows {\rm adj}$ & $\displaystyle\frac {N^2-1} {2N} \left|1\pm\frac {N} {2\lambda_f} \right|^2$ \\ \cline{2-3} 
\multirow{-2}{*}{$\SU(N)$} & $\tiny\Yvcentermath1\yng(1,1) \rightleftarrows \tiny\Yvcentermath1\yng(2)$ &        $\displaystyle\frac{N^3-N}8\left|1\pm \frac 1{\lambda_f}\right|^2$                                                         \\ \midrule
                           & $1\rightleftarrows {\rm adj}$                                          &                               $ \displaystyle (N-1) \left|1\pm\frac {N-2} {\lambda_f} \right|^2$                                      \\ \cline{2-3} 
\multirow{-2}{*}{$\SO(N)$} & ${\rm adj} \rightleftarrows \tiny\Yvcentermath1\yng(2)$                   &                    $\displaystyle\frac{N^3-N^2-4N+4}4\left|1\pm \frac 2 {\lambda_f}\right|^2$                                             \\ \midrule
\end{tabular}
\end{center}
\caption{\em  Group-theory factors for formation of a bound state  in the representation $R_f$ from an initial state in the representation $R_i$ and viceversa.
\label{tab:Ccolor2}}
\end{table}

Due to the attraction/repulsion of light mediators, the tree level cross-section is 
corrected by the Sommerfeld effect~\cite{hep-ph/0610249,0706.4071,astro-ph/0504621,1612.02825}
as $\sigma\approx S\times \sigma_0$, where 
$S$ encodes the effect of long-distance interactions that deflect the incoming fermion wave-function.  
The effect of SM vectors is known from the literature. 
We focus here on the effect of dark gluons. For $s$-wave annihilation 
\begin{equation}
S=\frac {2\pi \alpha_{\rm eff}/v_{\rm rel}}{1- e^{-{2\pi \alpha_{\rm eff}/v_{\rm rel}}}}
\end{equation}
where $\alpha_{\rm eff}$ is the effective coupling in each dark color channel as defined in eq.~(\ref{eq:coulombpot}).
The fermion bi-linears decompose in the representation of the dark-color group:
\begin{eqnsystem}{sys:Nmodel}\
\label{eq:Nmodeldec}
G_{\rm DC}=\SU(\ndc)&:& \quad\quad \ndc\otimes \bar{N}_{\rm DC} = \One\oplus {\rm adj},\qquad \ndc\otimes \ndc ={\tiny\Yvcentermath1\yng(2)  \oplus \tiny\Yvcentermath1\yng(1,1)}\\ 
G_{\rm DC}=\SO(\ndc )&:&\quad\quad
\ndc\otimes \ndc =\One\oplus  {\rm adj} \oplus {\tiny\Yvcentermath1\yng(2)}.
\end{eqnsystem}
The effective potential in each channel is given by eq.~\eqref{eq:coulombpot} with $\lambda_J=0$ and  $\lambda_I=\lambda_R$ where
\be
\begin{tabular}{c|c|c}
\multicolumn{3}{c}{\bf $G_{\rm DC}=\SU(\ndc)$}\\ \toprule
$R$ &$\lambda_{R}\times (2N)$& \hbox{bound states}   \\  \hline
1  &$\ndc^2-1$ & yes \\ 
adj &$-1$ & no  \\ 
{\tiny\Yvcentermath1\yng(2)} &$1-\ndc$ & no \\ 
{\tiny\Yvcentermath1\yng(1,1)} &$\ndc+1$ & yes  \\ 
\end{tabular}
\quad\quad\quad\quad
\centering
\label{my-label}
\begin{tabular}{c|c|c}
\multicolumn{3}{c}{\textbf{$G_{\rm DC}=\SO(\ndc)$}} \\ \toprule
$R$& $\lambda_R$ & \hbox{bound states} \\ \hline
1 & $ \ndc-1$ &  yes \\
\rm{adj} &1 & yes \\ 
{\tiny\Yvcentermath1\yng(2)} & $-1$ & no
\end{tabular}
\ee
Furthermore, two dark quarks can form a bound states emitting one vector.
A pair of dark quarks in the fundamental representation feels an attractive force in the
singlet and in the antisymmetric configuration. 
We adopt the results of~\cite{1702.01141} for the bound state formation cross sections.
For example, the cross section for forming the ground state, with quantum numbers $n=1$ and $\ell=0$, is
\be
(\sigma v_{\rm rel})_\text{bsf}
 =\frac 1 {N_F} \sigma _0   \lambda_i (\lambda_f \zeta)^5 \frac{2 S +1}{g_\chi^2} \, \frac{2^{11} \pi (1+\zeta ^2 \lambda_i^2) e^{-4 \zeta  \lambda_i  \text{arccot}(\zeta  \lambda_f )}}{3 (1+\zeta ^2 \lambda_f^2)^3
   \left(1-e^{-2 \pi  \zeta  \lambda_i}\right)}\times I_{R_i\rightarrow R_f}
\ee
where $\sigma_0\equiv \pi\adc^2/\MDM^2$, $\zeta\equiv \adc/v_{\rm rel}$ and  $\lambda_{i,f}$ are the effective strength of the coupling $\alpha_{\rm eff}\equiv \lambda_I \adc$ of the initial and final state channels respectively,
in two-body representation $R_i$ and $R_f$.
The  $I_{R_i\rightarrow R_f}$ factors 
encode the group theory structure and are listed in table~\ref{tab:Ccolor2}.

\begin{table}
{\small
\begin{tabular}{ccccc|cccc|c|c}
\hbox{Name}& $I$ & $S$ & $n$ & $\ell$ &$\Gamma_{\rm ann}/\MDM$&$N_{\rm DC}=3$ & $N_{\rm DC}=4$& $N_{\rm DC}=5$& $\Gamma_{\rm dec}/\MDM$ & \hbox{Prod. from}\\  \toprule
$1s_1^-$ & 1 & 0 & 1 & 0 & & $(8/6)^3\alpha_{\rm DC}^5$  & $(15/8)^4\adc^5$ &$3(24/10)^3\adc^5$&  0 & $p_{\rm adj}$  \\
$1s_1^+$ & 1 & 1 & 1 & 0 &$\frac{5(\pi^2-9)}{\pi}\times$ & $ 2^6\alpha_{\rm DC}^6/3^7$ &$15^3\adc^6/2^{14}$ &$3^3(2/5)^6\adc^6/7 $& 0 & $p_{\rm adj}$  \\ 
$1s_{\scalebox{.19}{\yng(1,1)}}$ & \scalebox{0.33}{\yng(1,1)}  &  1 & 1 & 0  & &0 & 0&0 & 0& $p_{\scalebox{.19}{\yng(2)}}$ \\  \midrule
$2s_1^-$ &1 & 0 & 2 & 0 & &$(8/12)^3\alpha_{\rm DC}^5$  &$15^4\adc^5/8^5$ & $3(24/20)^3\adc^5$&  ${\cal O}(\alpha_{\rm DC}^6)$  &$ p_{\rm adj}$ \\
$2s_1^+$ &1 & 1 & 2 & 0& $\frac{5(\pi^2-9)}{\pi}\times$ &$2^3\adc^6/3^7$  &$15^3\adc^6/2^{17}$ &$6^3/7(\adc/5)^6$ &  ${\cal O}(\alpha_{\rm DC}^6)$  &$ p_{\rm adj}$ \\
$2s_{\scalebox{.2}{\yng(1,1)}}$ & \scalebox{0.32}{\yng(1,1)} & 1 & 2 & 0  & &  ${\cal O}(\alpha_{\rm DC}^6)$&   ${\cal O}(\alpha_{\rm DC}^6)$&   ${\cal O}(\alpha_{\rm DC}^6)$& ${\cal O}(\alpha_{\rm DC}^6)$ & $p_{\scalebox{.19}{\yng(2)}}$ \\  \midrule
$2p_1^-$ & 1 & 0 & 2 & 1  & & ${\cal O}( \alpha_{\rm DC}^7)$ & ${\cal O}( \alpha_{\rm DC}^7)$& ${\cal O}( \alpha_{\rm DC}^7)$& $ {\cal O}(\alpha_{\rm DC}^6)$  & $s_{\rm adj}$ \\
$2p_1^+$ & 1 & 1 & 2 & 1 & & ${\cal O}( \alpha_{\rm DC}^7)$ & ${\cal O}( \alpha_{\rm DC}^7)$& ${\cal O}( \alpha_{\rm DC}^7)$& $ {\cal O}(\alpha_{\rm DC}^6)$  & $s_{\rm adj}$ \\
$2p_{\scalebox{.2}{\yng(1,1)}}$ & \scalebox{0.32}{\yng(1,1)} & 0 & 2 & 1  & & ${\cal O}( \alpha_{\rm DC}^7)$ & ${\cal O}( \alpha_{\rm DC}^7)$& ${\cal O}( \alpha_{\rm DC}^7)$& $ {\cal O}(\alpha_{\rm DC}^6)$ & $s_{\scalebox{.2}{\yng(2)}}$ \\ \midrule
\end{tabular}}
\caption{\emph{Summary of perturbative di-quark bound states  in $\SU(N)$ models.}}
\label{tab:NBS}
\end{table}

\subsection{Model with $G_{\rm DC} = \SU(3)$ and singlet dark quark}\label{sec:SU3N}
We first consider the model where the dark quark $\Q$ in the fundamental of $\SU(\ndc)$ is a singlet under the SM.
We assume that  extra unspecified heavier dark quarks with SM  charges couple the dark sector with the SM sector, such that glue-balls decay into SM particles.
The $s$-wave $\Q\bar\Q$ annihilation cross-section  into dark gluons is
\be
\langle\sigma v_{\rm rel} \rangle=\frac{N_{\rm DC}^4- 3 N_{\rm DC}^2+2}{16N_{\rm DC}^3}\left(\frac 2 {N_{\rm DC}^2-2}S_{{1}}+\frac{N_{\rm DC}^2-4}{N_{\rm DC}^2-2} S_{\rm adj}\right) \times \frac{\pi \alpha_{\rm DC}^2}{\MDM^2}
\ee
where $S_{1}$ and $S_{\rm adj}$ are the Sommerfeld factors for the singlet (attractive) and adjoint (repulsive) channels.

Let us consider the bound states.
The $\SU(\ndc)$ interactions give two attractive configurations that can support bound states: 
the singlet contained in $\Q\otimes \bar{\Q}$ and the anti-symmetric configuration  in $\Q\otimes \Q$. 
The former is unstable and  gives a contribution to the effective annihilation cross section, see Appendix; 
the latter is stable and could give rise to dark-recombination at low temperatures ($T\lesssim \alpha_{\rm DC}^2 \MDM$). 
The unstable bound state is made of Dirac particles so it exists for any choice of quantum numbers $n,\ell,s$.  
The stable bound state   is made of identical particles, so that a fully anti-symmetric wave-function 
implies that it must have spin $1$ in $s$-wave and spin-0 in $p$-wave. 
Moreover it can only be produced from an initial state in the symmetric configuration. 
The main bound states together with their key properties are summarized in table \ref{tab:NBS}.

\begin{figure}[t!] 
\begin{center}
\includegraphics[width=0.47\textwidth]{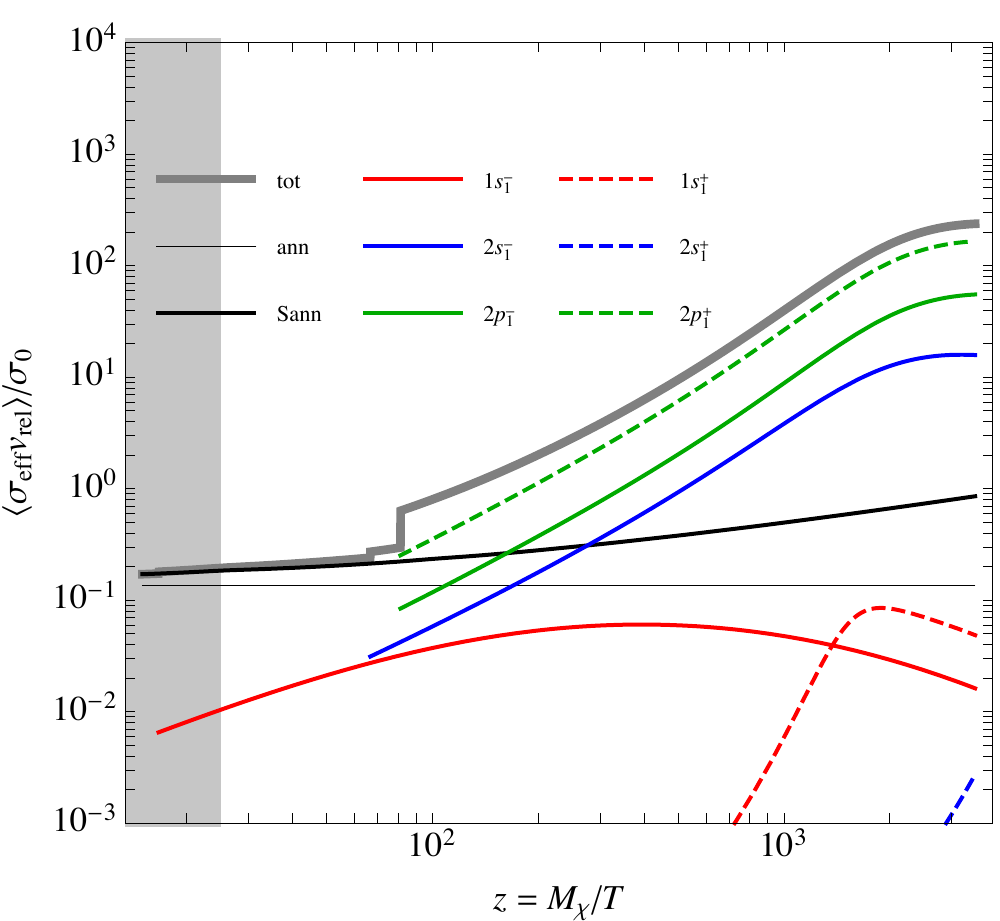}\qquad
\includegraphics[width=0.42\textwidth]{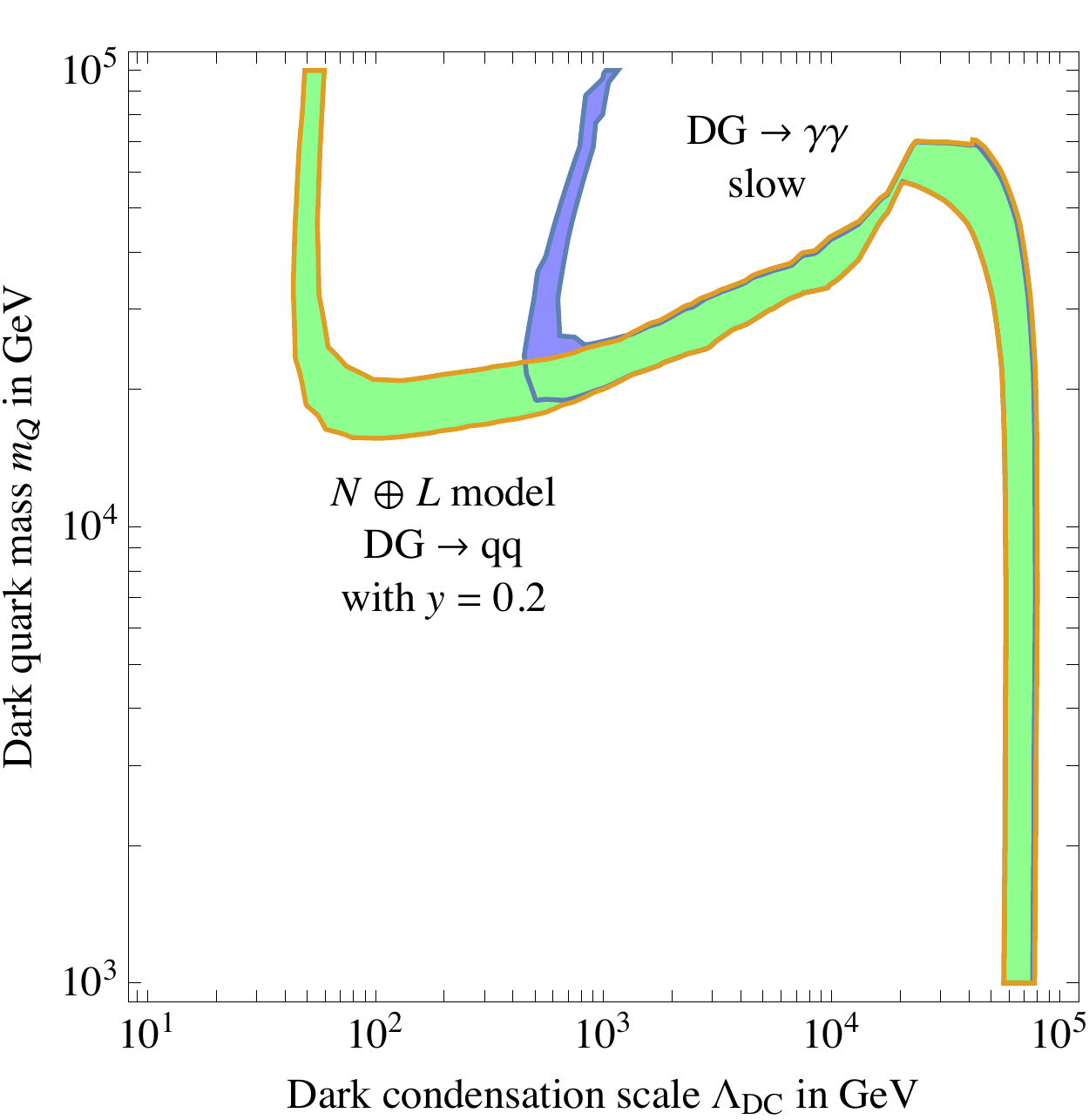}
\end{center}
\caption{\emph{ Model with $G_{\rm DC} = \SU(3)$ and a dark quark neutral under $G_{\rm SM}$. 
{\bf Left}: thermally averaged  cross sections for annihilation
and for  bound states formation, assuming  $ m_\Q = 10 \TeV$ and $ \adc = 0.1$ ($\LDC\approx 30 \GeV$).
{\bf Right}: region where  dark baryons reproduce the DM cosmological abundance. A recombination fraction $\wp_\B=0.4$ is assumed.}}
\label{fig:relicN}
\end{figure}
\medskip

If  dark confinement happens after  freeze-out,
the thermal relic abundance of DM is obtained by first solving the Boltzmann equations for the elementary dark  quarks and their perturbative bound states.  
Table~\ref{tab:NBS} implies that the bound states are produced from a repulsive initial state. 
This suppresses the production of stable and unstable di-quark bound states at late times, where the kinetic energy is insufficient to overcome the repulsion. As a consequence, we find that  the thermal relic abundance is mostly due to perturbative annihilations  boosted by the Sommerfeld enhancement, 
and by di-quark bound state production at earlier times.
At $T\sim \LDC$ confinement occurs in the dark sector, and a fraction of the dark quarks is converted into dark baryons. The dark baryons can undergo recombination annihilations, which have large cross sections, leading to a late-time dark matter depletion. 

When dark confinement takes place before  freeze-out, annihilations are still governed by the constituent cross section, provided that
the typical velocities at freeze-out are large enough.  At lower velocities,  the larger recombination cross section produces a late-time dark matter depletion. 

Taking all these effects into account, fig.~\ref{fig:relicN} shows an estimate of the
parameter region where the thermal relic abundance of dark baryons matches the cosmological DM abundance.


\begin{figure}[t]
\begin{center}
\includegraphics[width=.45\textwidth]{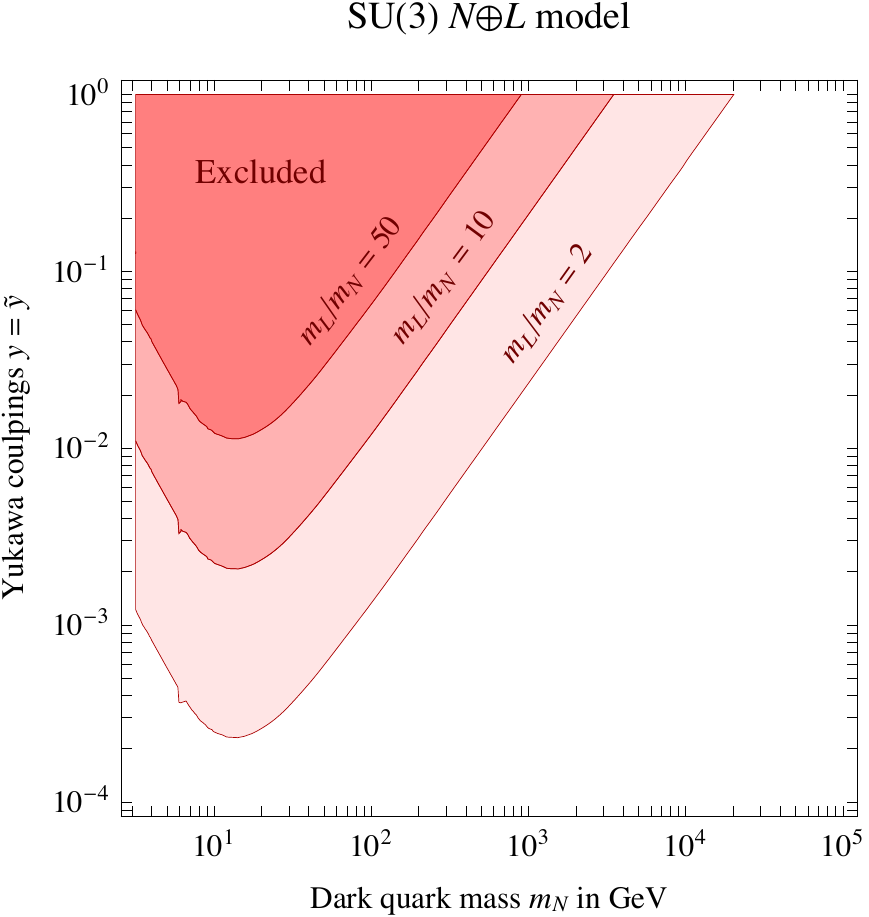}\qquad
\includegraphics[width=.45\textwidth]{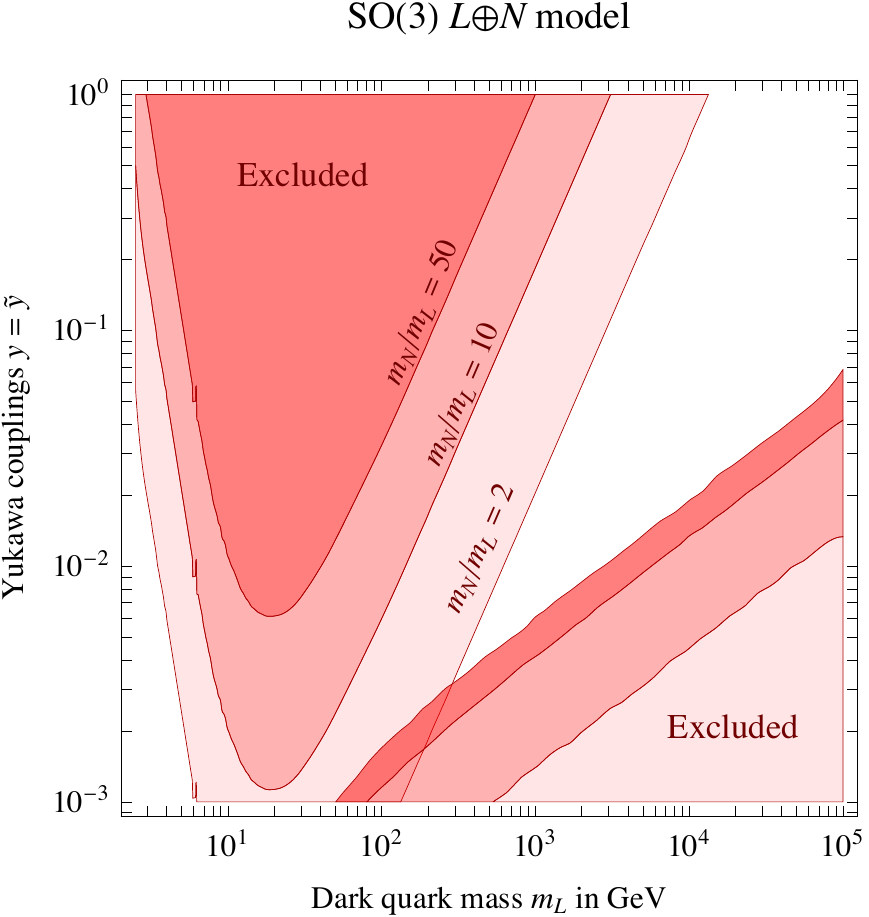}
\caption{\em Direct detection bounds, assuming dark quarks  $\Q=N\oplus L$ with Yukawa couplings to the Higgs.
{\bf Left:} we consider the $\SU(3)$ model with $m_N < m_L$.
{\bf Right:} we consider the $\SO(3)$ model with $m_L< m_N$, such that a large enough Yukawa coupling
is needed in order to suppress $Z$-mediated inelastic scatterings.}
\label{fig:DDBound}
\end{center}
\end{figure}

\medskip

A  dark quark $\Q$ singlet under the SM can interact with the SM sector through heavier mediators. The most interesting possibility 
is realised adding a vector-like dark quark $L$, allowing to write Yukawa couplings with the SM Higgs
\be\label{eq:SUYukawas}
-\La =  m_L L L^c + m_N N N^c+y\, L H N^c+\tilde y\, L^c H^\dagger N+\hbox{h.c.}
\ee
As explained in section \ref{sec:dd}, after electro-weak symmetry breaking the singlets mix with the neutral component of the doublet  generating an effective coupling to the $Z$ and to the Higgs.
Denoting with $U_L$ and $U_R$ the rotation matrices to the mass eigenstate basis, the coupling to $Z$ is
\begin{equation}
\frac {g_2}{2\cos\theta_{\rm W}} Z_\mu \left(\bar{N}_i (U_L^\dagger)_{2i} \bar{\sigma}^\mu  (U_L)_{2j} N_j - \bar{N}_i^c (U_R^\dagger)_{2i} \bar{\sigma}^\mu  (U_R)_{2j} N^c_j \right)\,.
\end{equation}
For real Yukawa couplings (no CP violation) the $U_{L,R}$ are $\SO(2)$ matrices with rotation angle
\be
\tan 2\theta_L =\frac{2\sqrt2v\left(m_L\tilde y+m_Ny\right)}{2m_L^2-2m_N^2+(yv)^2-(\tilde yv)^2}
\ee
for $U_L$ and similarly for $U_R$. The light singlet dark quark $N$ acquires the coupling
\begin{equation}
\frac {g}{2\cos\theta_{\rm W}} Z_\mu \left(( s_L^2+ s_R^2) \bar{N} \gamma_\mu N-( s_L^2- s_R^2) \bar{N} \gamma_\mu\gamma_5 N\right)\,.
\end{equation}
Bounds from Higgs-mediated interactions are typically weaker and have a different dependence on the mixings, namely 
\be
\frac{h}{\sqrt2}\left(\tilde y\,c_Ls_R+y\,c_Rs_L\right) \bar NN\,.
\ee 
Fig.\ref{fig:DDBound} shows the bounds on the Yukawa coupling $y$,
once we combine Higgs-mediated and $Z$-mediated effects.
Experiments are sensitive even to heavy and weakly mixed fermions.

\subsection{Model with $G_{\rm DC}=\SU(3)$ and dark quark triplet under $\SU(2)_L$}\label{sec:SU3V}

\begin{figure}[t!] 
\begin{center}
\includegraphics[width=0.45\textwidth]{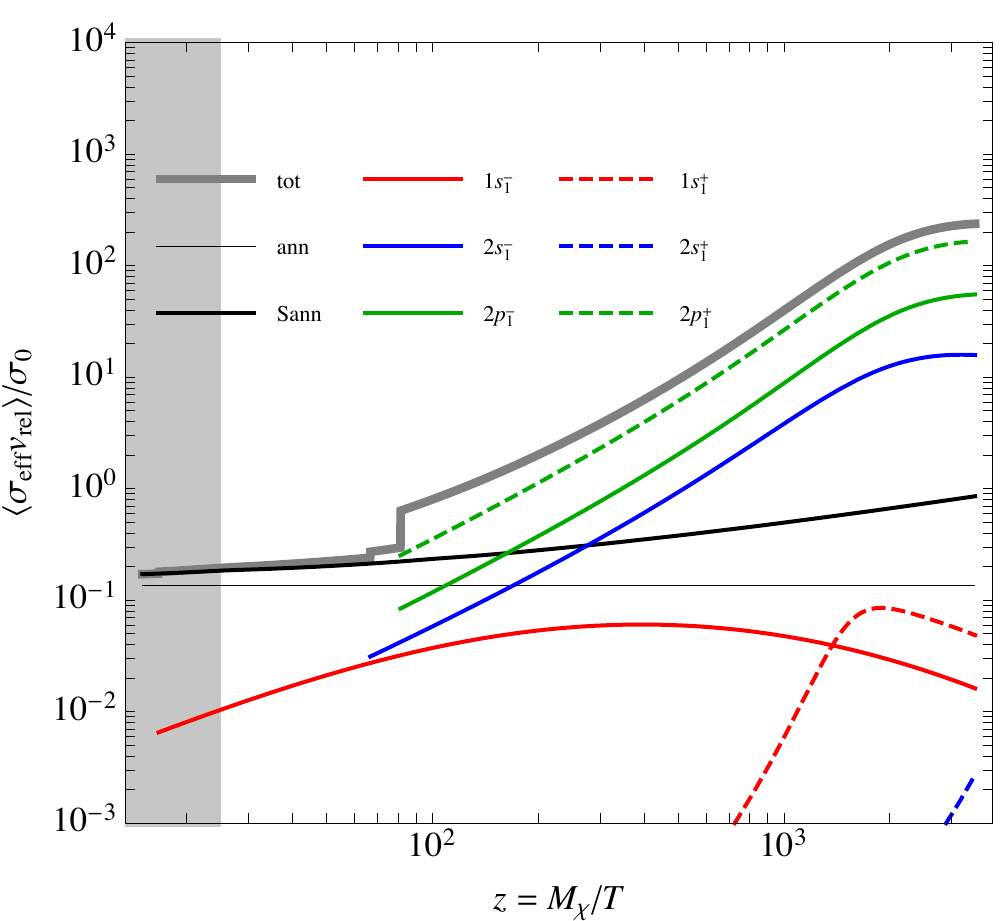}\qquad
\includegraphics[width=0.41\textwidth]{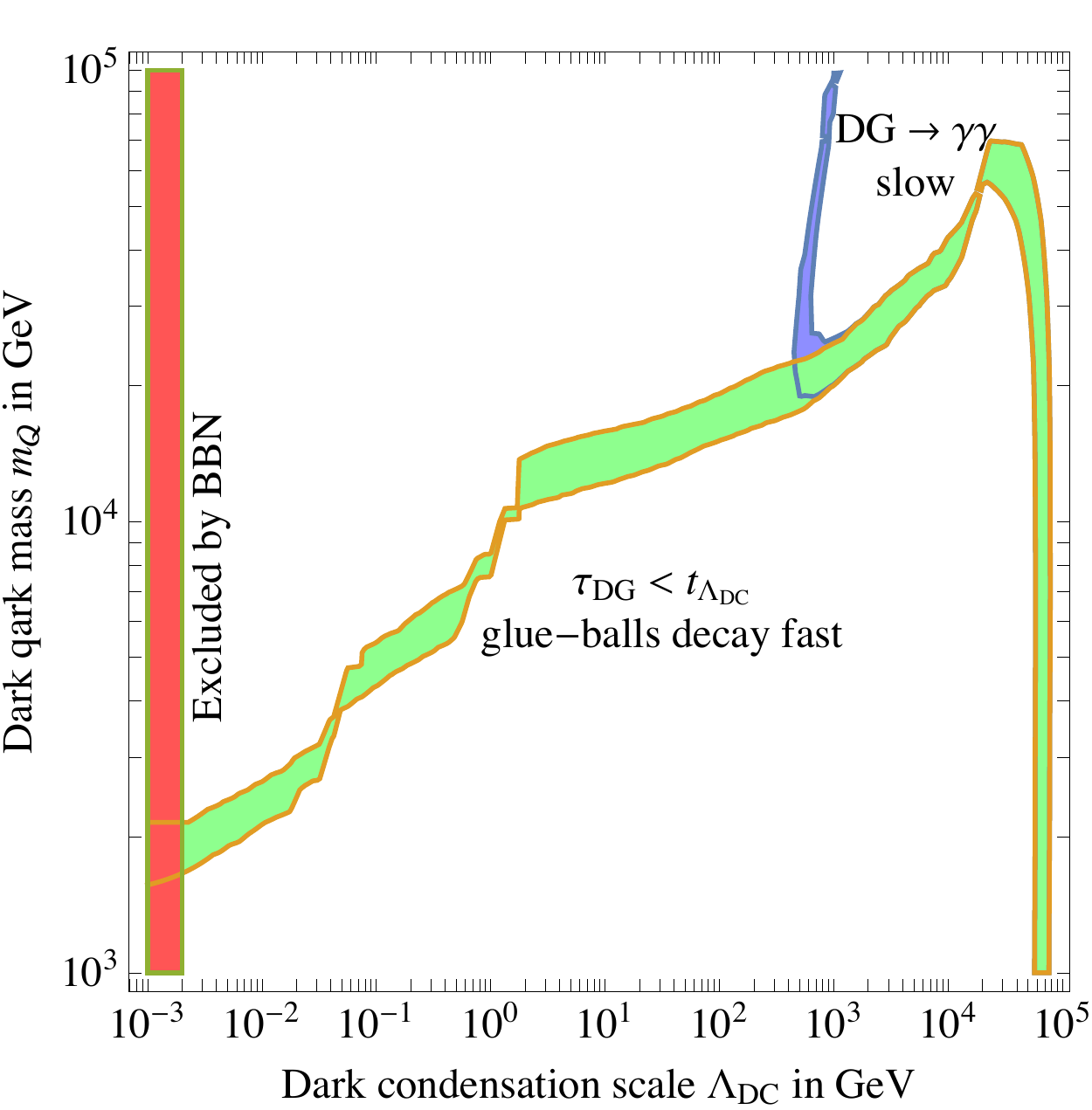}
\end{center}
\caption{\emph{Model with $G_{\rm DC} = \SU(3)$ and with a dark quark tripled under $\SU(2)_L$. 
{\bf Left:} thermally averaged  cross sections for annihilation
and for  bound states formation, assuming  $ m_\Q = 10 \TeV$ and $ \adc = 0.07$ ($\LDC\approx 3$ GeV).
{\bf Right:} region where  dark baryons reproduce the DM cosmological abundance.}}
\label{fig:relicV}
\end{figure}

We next consider the $G_{\rm TC}=\SU(\ndc)$ model  with dark quarks in a triplet ($V$) of $\SU(2)_L$. 
Requiring no sub-Planckian Landau poles selects $N_{\rm DC}=3$. 
We  compute in terms of $\SU(2)_L$ multiplets,
neglecting the 165 MeV electro-weak splitting between charged and neutral components.
SM gauge interactions keep the dark sector in thermal equilibrium with the SM sector.
Pairs of dark quarks decompose as
\beq
\Q\otimes \overline \Q = (1 \oplus 8 , 1 \oplus3 \oplus5) ,\qquad
\Q\otimes \Q =  (\overline{3}\oplus6, 1\oplus3 \oplus5) .
\eeq
The annihilation cross-section among dark quarks is
\begin{equation}
\langle\sigma v_{\rm rel} \rangle=\left(\frac 7 {162}\frac{\pi \alpha_{\rm DC}^2}{\MDM^2}+\frac 8 {27}\frac{\pi \alpha_{\rm DC} \alpha_2}{\MDM^2}+\frac {37}{72}\frac{\pi \alpha_2^2}{\MDM^2}\right)\left(\frac 2 7 S_{1}+\frac 5 7 S_{8}\right)\,.
\end{equation}
where $\lambda_1= 4/3$ and $\lambda_8=-1/6$ are the effective strengths of the Sommerfeld factors for the singlet and octet channels.  
For $\adc <3 \alpha_2$  the annihilation cross-section is dominated by the SM interactions. 

In the absence of confinement the desired DM relic abundance is obtained for $m_\Q \approx 2.5{\rm TeV}/\sqrt{2N_{\rm DC}} $;
such a model is however only allowed for $\adc\lesssim 10^{-8}$~\cite{1505.03542}. 
We assume that dark interactions dominate or are comparable with the SM ones.

Neglecting SM interactions, the meson bound states are listed in table \ref{tab:NBS}.
Each bound state has 9 components and decomposes as $1\oplus 3 \oplus 5$ under $\SU(2)_L$.
The singlet and quintuplet are symmetric under $\SU(2)_L$, so
they have the same spin as in the previous case $\Q=N$ listed in table \ref{tab:NBS}. 
The triplet have the opposite spin, being anti-symmetric under $\SU(2)_L$.

The lightest baryons have spin 1/2 and lie in the adjoint representation of  flavour $\SU(3)_F$,
and split  as $8=3\oplus5$ taking $\SU(2)_L$ gauge interactions into account, such that the triplet is lighter than the quintuplet.

Predictions for direct detection are then the same as for any fermion weak triplet (such as wino~\cite{0903.3381}):
$\sigma_{\rm SI}$  lies above the the neutrino background for $m_\B \lesssim 15$ TeV. 
Constraints on Yukawa couplings with heavier 
dark quarks are similar to those discussed in the $N \oplus L $ model.

The annihilation cross-section relevant for indirect detection a few orders of magnitude above the
canonical thermal value $3~10^{-26}\cm^3/\sec$, being dominated by  long-range rearrangement processes 
as discussed around eq.~(\ref{eq:indirectxsec}); presumably without extra Sommerfeld enhancement.
Present bounds are shown in fig.~\ref{fig:IndirectV}, as a function of the dark glue-ball mass which
controls the energy spectrum of final-state particles. 
We combine searches for diffuse gamma rays from the {\sc FermiLAT} satellite and from the ground based  H.E.S.S.\
observatory   The FermiLAT limits are more relevant in the case of light glue-balls decaying into photons;
 the H.E.S.S.\ limits are sensitive to the cascade photons resulting from $W$ boson decays in case of heavy glue-balls. The sensitivity of the  photon searches   strongly depends on the number of steps in the dark hadronization cascade and is thus rather uncertain. The limits coming from annihilation into $WW$ are more robust.  We also show the limits from CMB energy injection are shown, which have smaller astrophysical and theoretical uncertainties. 

\subsection{Models with $G_{\rm DC}=\SO(\ndc)$}
As discussed in section~\ref{models}, models with dark gauge group $\SO(\ndc)$ give rise to Majorana DM, allowing for
lightest dark quakrs in more general representations under $G_{\rm SM}$.
The annihilation cross-section of fermions in the fundamental of $\SO(\ndc)$  into dark gluons is
\be
\langle\sigma v_{\rm rel} \rangle=\frac{N_{\rm DC}^2(N_{\rm DC}-1)}{2}\left(\frac 4 {N_{\rm DC}^2} S_{\bold{1}}+\frac {N_{\rm DC}^2-4} {N_{\rm DC}^2} S_{\tiny\yng(2)}\right) \times \frac{\pi \alpha_{\rm DC}^2}{\MDM^2}
\ee
where $S_{\bold 1}$ and $S_{\rm adj}$ are the Sommerfeld factors for the singlet and adjoint channel respectively.
As a simple example we consider the model with a singlet $N$ and a doublet $L$,
\be
\label{eq:SUYukawas}
-\La =  m_L L L^c + \frac{m_N}2 N^2+y\, L H N+\tilde y\, L^c H^\dagger N+\hbox{h.c.}
\ee
Differently from the singlet model in section \ref{sec:SU3N}, $N$ and $N^c$ are the same particle.
The mass matrix of the neutral states is
\be
\label{eq:SOYukawas}
\La\supset\frac{1}{2} \left(N_1,N_2, N_3\right)\left(\begin{array}{ccc}0 & m_L & vy/\sqrt2\\ m_L & 0 & v\tilde y/\sqrt2\\vy/\sqrt2 & v\tilde y/\sqrt2 & m_N\end{array}\right)\left(\begin{array}{c}N_1\\N_2\\N_3\end{array}\right)+\hbox{h.c.}
\ee 
where the Weyl fermions
$N_1$ and $N_2$ are the neutral components of $L$ and $L^c$ and $N_3\equiv N$. 
The  mass matrix can be diagonalised as $M_{\rm diag}= U^T M U$, where, at leading order in the Yukawa couplings
\begin{equation}
U= \left(
\begin{array}{ccc} 
\frac 1 {\sqrt{2}} & \frac i {\sqrt{2}} & -\frac {\tilde{y}v}{\sqrt{2}(m_L-m_N)}\\ 
\frac 1 {\sqrt{2}} & -\frac i {\sqrt{2}} & -\frac { y v}{\sqrt{2}(m_L-m_N)}\\
\frac {v (y +\tilde{y})}{2(m_L-m_N)} &   \frac {i(y-\tilde{y})v}{2(m_L-m_N)}&1
\end{array}\right)\,.
\end{equation}
The gauge coupling to the $Z$ in the flavor basis are $Q_Z={\rm diag}(1/2,- 1/2, 0)$. Rotating to the mass basis we obtain the couplings of the mass eigenstates to the $Z$,
\begin{equation}\label{eq:Zcoupling}
g_{ij}\equiv \left(U^\dagger Q_Z U\right)_{ij} =
\left(\begin{array}{ccc} 
0 & \frac i {2} & \frac {(y-\tilde{y})v}{4(m_L-m_N)}\\ 
-\frac i 2 & 0 & i\frac {(y+\tilde{y})v}{4(m_L-m_N)}\\
\frac {(y^*-\tilde{y}^*)v}{4(m_L-m_N)} & -i\frac {(y^*+\tilde{y}^*)v}{4(m_L-m_N)} & \frac {-(|y|^2+|\tilde{y}|^2)v^2}{4(m_L-m_N)^2}
\end{array}\right)_{ij}\,.
\end{equation}
Because the mass eigenstates are Weyl fermions, the diagonal couplings of DM to the $Z$ are purely axial. This can be made manifest converting to Majorana notation  $\Psi_M\equiv  (N, \bar N)/\sqrt{2}$ such that $\bar{\Psi}^i_M \gamma^\mu \Psi_M^i$ vanishes identically. 
In this basis one finds
\begin{equation}
\frac{g_2}{\cos \theta_{\rm W}}Z_\mu \left[ a_{ij} \bar  \Psi_M^i \gamma^\mu  \gamma^5\Psi_M^j+i  v_{ij} \bar  \Psi_M^i \gamma^\mu  \Psi_M^j \right]
\end{equation}
where $a_{ij}=-{\rm Re}\, g_{ij}$ and $v_{ij}={\rm Im}\, g_{ij}$. From eq.~\eqref{eq:Zcoupling} we see that the only non vanishing terms are of the form $\bar\Psi^i_M\gamma^\mu\gamma^5\Psi_M^i$ and $\bar\Psi^i_M\gamma^\mu\Psi_M^j$ with $i\ne j$. The first interaction gives rise to spin-dependent interactions suppressed by the mixing with the heavier states,
which are below the sensitivity of present experiments.
The second interaction produces inelastic scattering between states with a mass splitting of order $\Delta m\sim y^2v^2/(m_N-m_L)$. 

The Higgs-mediated contribution to direct detection is similar to $\SU(\ndc)$ models. 
Writing $\La _H= y_{ij} h N^i N^j/\sqrt{2}+\hbox{h.c.}$ one finds
\begin{equation}
y_{ij} \equiv \left(U^T \frac {\partial M(h)}{\partial h} U\right)_{ij}=
\left(\begin{array}{ccc} 
\frac {(y+\tilde y)^2v}{2 (m_L-m_N)} & \frac {i(y^2-\tilde{y}^2) v}{2(m_L-m_N)} & \frac {(y+\tilde{y})}{2}\\ 
 \frac {i(y^2-\tilde{y}^2) v}{2(m_L-m_N)} & -\frac {(y-\tilde y)^2v}{2 (m_L-m_N)}  & \frac {i(y-\tilde{y})}{2}\\
\frac {(y+\tilde{y})}{2} &\frac {i(y-\tilde{y})}{2} & - \frac {2 y \tilde yv}{m_L-m_N}
\end{array}\right)_{ij}\,.
\end{equation}
Fig.~\ref{fig:DDBound} illustrates the present bounds on the Yukawa couplings.

\section{Conclusions}\label{Conclusions}
We studied fundamental theories of Dark Matter as baryons made of a dark quark $\Q$ with mass $m_\Q$,
charged under a dark gauge group $\SU(\ndc)$ or $\SO(\ndc)$ that becomes strong at a scale $\LDC$.
The main options for the gauge quantum numbers of $\Q$ are: either neutral or charged under the SM gauge group.
DM is stable because dark baryon number is accidentally conserved, analogously to the proton in the SM.

In past works we studied the possibility that $\Q$ is lighter than the
dark condensation scale $\LDC$, finding
that the DM cosmological abundance was reproduced as a thermal relic for $\LDC\sim 100\TeV$,
which saturates the perturbative unitarity bound on DM annihilations.
In this work we explored the opposite situation:
this simple generalization leads to unusual and non-trivial  DM phenomenology.

The dark confinement scale $\LDC$ can be as low as $0.1\GeV$, giving rise to unstable dark glue-balls
with mass $\mDG \sim 7 \LDC$ as lightest dark particles.
Dark glue-balls decay into lighter SM particles, and can be searched for in low-energy experiments.

In cosmology, dark quarks freeze-out as usual at $T\sim m_\Q/25$.
DM can be lighter than $100\TeV$ because the dark gauge coupling $\adc$ is perturbative, when renormalized at this energy.
However, a second stage of cosmological history contributes to determining the DM relic abundance:
after a first-order phase transition at $T \sim \LDC$ (that can lead to gravitational waves)
the dark quarks must bind into objects neutral under dark color:
a fraction of dark quarks forms dark mesons, that decay, 
the rest binds into stable dark baryons $\B$ that survive as DM.
We estimated this fraction in a geometric model of dark hadronization, that takes into account that dark strings do not break.
As a consequence the annihilation cross section among dark quarks, $\sigma_{\Q\bar\Q} v_{\rm rel} \sim \pi \adc^2/m_\Q^2$ can be smaller than the standard
cosmological value, $3~10^{-26}\cm^3/\sec$.

More importantly, the annihilation cross section among dark baryons, $\sigma_{\B\bar\B} v_{\rm rel} \sim 1/\adc m_\Q^2$, is typically much larger than $\sigma_{\Q\bar\Q}$, being enhanced by a negative power of $\adc$.
This happens because annihilation can proceed through an atomic-physics process, recombination: 
at low enough energy a dark quark $\Q$ in a dark baryon $\B$ can recombine forming a meson with a $\bar\Q$ in a $\bar B$;
afterwards the meson decays through the usual particle-physics $\Q\bar\Q$ annihilation.
If $m_\Q \gg \LDC$ the bound state $\B$ is dominated by the Coulombian part of the potential,
and this is similar to recombination occurring in hydrogen anti-hydrogen scattering.
We computed the binding energies of dark baryons and mesons by means of a variational method, finding that recombination is kinematically allowed.
If instead $m_\Q\circa{>}\LDC$ the confining part of the potential is relevant, and the process can be seen
as the crossing of dark strings (flux tubes of the dark color interaction).
In cosmology, the large $\sigma_{\B\bar\B}\gg \sigma_{\Q\bar\Q}$ leads to extra dilution of the DM density.
In astrophysics, it leads to large signals for indirect DM detection.
Dark mesons decay into dark glue-balls: depending on the model their decays might be dominated by gauge couplings
(producing photons) or by Yukawa couplings (producing leptons, which can provide a DM interpretation of
the $e^+$ excess observed by PAMELA and AMS~\cite{PAMELA})).

Cosmological evolution leads to the formation of excited dark baryons,
which quickly decay into glue-balls proved that their excitation energy $\Delta E_B \sim \adc^2 m_\Q$
is larger than $\mDG\sim 7\LDC$.  Otherwise, de-excitation can be slow, 
proceeding through off-shell dark glue-balls,
giving rise to a novel phenomenon:
dark matter that emits either $\beta$ or $\gamma$ radioactivity (again depending on whether
gauge couplings of Yukawa couplings dominate).
It would be interesting to explore whether radio-active DM can alleviate the core/cusp and missing-satellite issue of cold DM.

Finally, we studied the direct-detection and collider phenomenology of models where DM is made of heavy baryons.   
Heavy dark quarks  can be produced at colliders, manifesting as narrow spin-0 or spin-1 resonances and producing effects in SM precision observables. Current bounds are consistent with a lightest dark quark charged under the SM heavier than 1-2 TeV.

\small

\subsubsection*{Acknowledgments}
We thank Gennaro Corcella, Francesco Becattini, 
Massimo Porrati and Gabriele Veneziano
for useful discussions.
This work was supported by the ERC advanced grant 669668 (NEO-NAT). 
JS and AM thank the CERN theory division for the hospitality during the final state of this project.

\appendix

\section{Boltzmann equations}\label{sec:AppA}
We use the Boltzmann equations for dark quarks and for their bound states written in~\cite{1702.01141}.
For $T> \LDC$ the cosmological evolution of the abundance of dark quarks $\Q$  and their bound states $I$ is described by a set of coupled Boltzmann equations,
\begin{align} \label{eq:YDM}
&sHz \frac{dY_{\mathcal{Q}}}{dz} = -2\gamma_{\rm ann}
\bigg[\frac{Y_{\mathcal{Q}}^2}{Y_{\mathcal{Q},\rm{eq}}^2 }-1\bigg]-2\sum_{I}
\gamma_I  \bigg[\frac{Y_{\mathcal{Q}}^2}{Y_{\mathcal{Q},\rm{eq}}^2} -\frac{Y_I}{Y_{I, \rm{eq}}}\bigg] \\
& sHz\frac{dY_I}{dz} =n_I^{\rm eq}\bigg\{
 \med{ \Gamma_{I\rm break} } 
\bigg[\frac{Y_{\mathcal{Q}}^2}{Y_{\mathcal{Q},\rm{eq}}^{2}} - \frac{Y_I}{Y_{I,\rm eq}}\bigg]
+\med{\Gamma_{I\rm ann}} 
\bigg[1-\frac{Y_I}{Y_{I,\rm{eq}}}  \bigg]+\sum_J
\med{\Gamma_{I\to J}} 
\bigg[\frac{Y_J}{Y_{J\rm, eq}}-\frac{Y_I}{Y_{I,\rm eq}}  \bigg]  \bigg\}.\nonumber
\end{align}
where $Y_{\mathcal{Q},I} = n_{\mathcal{Q},I}/s$ with $s$ the entropy density, $z = \mQ/T$. We define as $n^{\rm eq}$ and $Y^{\rm eq}$ the thermal equilibrium value of $n$ and $Y$ respectively and $\gamma$ is the space-time density of interactions in thermal equilibrium, as defined in~\cite{0903.3381}. The first term describes $\Q\bar\Q$ annihilations to SM particles; the second term describes formation of the bound state identified by the index $I$ that collectively denotes its various quantum numbers:
angular momentum, spin, gauge group representation, etc. 

The effect of rapidly unstable bound states can be encoded in an effective annihilation rate, $\gamma_{\rm ann}^{\rm eff}$,
that substitutes $\gamma_{\rm ann}$, such that their Boltzmann equations can be dropped.
In this way, \cite{1702.01141} managed to obtain a single Boltzmann equation.
However, the present study contains a new feature: some bound states (such as $\Q\Q$) do not decay,
and can only be formed or broken by interactions.
We then need to separately evolve the Boltzmann equations for their abundances. 
We define 
$\gamma_{\rm bsf-stable}=\sum_{I} \gamma_I$ with the sum running over the unstable  bound states,
and similarly for the stable ones.
In the non-relativistic limit the space-time densities $\gamma$ get approximated as
\beq 2\gamma \stackrel{T\ll \MDM}{\simeq} (n_{\mathcal{Q}}^{\rm eq})^2 \med{\sigma v_{\rm rel}}\eeq
such that the Boltzmann equations simplify to
\be
\left\{\begin{split}
& \frac{1}{\lambda} \frac{d Y_\mathcal{Q}}{d z }  = - \frac{S_{\rm eff-unstable} }{z^2} \left( Y_\mathcal{Q}^2 - Y_{\mathcal{Q},\rm{eq}}^2  \right) -  \frac{ S_{I , \rm bsf} }{z^2} \left( Y_\mathcal{Q}^2 - Y_I \frac{Y_{\mathcal{Q}, \rm eq}^2}{Y_{I, \rm eq}}\right) \\ 
& \frac{1}{\lambda} \frac{d Y_I}{d z}  =  \frac{S_{I , \rm bsf} }{z^2} \left( Y_\mathcal{Q}^2 - Y_I \frac{Y_{\mathcal{Q}, \rm eq}^2}{Y_{I, \rm eq}}\right) \,,
\end{split}\right.
\ee
where we introduced the dimension-less factors $S_{\rm eff-unstable} = S_{\rm ann} + S_{\rm bsf-unstable}$ and
\be
 S_{X}(z) = \frac{\med{\sigma_X v_{\rm rel}}}{\sigma_0} ,
\qquad
 \lambda = \left.\frac{\sigma_0 s}{H}\right|_{T=\MDM} = 
 \sqrt{\frac{g_{\rm SM}\pi}{45}} \sigma_0 M_{\rm Pl} \MDM\,.
\ee
Here $g_{\rm SM}$ is the number of degrees of freedom in thermal equilibrium at $T=\MDM$
($g_{\rm SM}=106.75$ at $T\gg M_Z$).


Stable bound states $I$ are kept into thermal equilibrium by fast dark gauge interactions,
so that they decouple at a $z_I$ much later than DM freeze-out, that occurs at $z_f \sim 25$.
Thereby for $z  \ll z_I$ we obtain a single Bolztmann equation
\be
\frac{1}{\lambda} \frac{d Y_\mathcal{Q}}{d z }  = - \frac{S_{\rm eff} }{z^2} \left( Y_\mathcal{Q}^2 - Y_{\mathcal{Q},\rm{eq}}^2  \right) ,\qquad
S_{\rm eff} = S_{\rm ann}+S_{\rm bsf-unstable} + S_{\rm bsf-stable}
\ee
approximatively solved by~\cite{1702.01141}
\begin{align}\label{eq:Yapprox}
Y_\mathcal{Q}(z) = \frac{1}{\lambda}\left( \int_{z_f}^z \frac{S_{\rm eff} (z)}{z^2} dz + \frac{S_{\rm eff}  (z_f)}{z_f^2} \right)^{-1}\,.
\end{align}
We now compute $z_I$, showing that it is so large that later annihilations are negligible.
The value of $z_I$ is needed to estimate the fraction of dark quarks bound in stable states.

Assuming that  $ \sfrac{d Y_I}{d z }  \approx 0 $ is violated at  $z_I$ so large
that annihilation processes are negligible,  we  have  $Y_\mathcal{Q}(z) + 2 Y_I(z) =  Y_\mathcal{Q}(z_I) =Y_c$ at temperatures 
$z>z_I$
at which the stable bound states are no longer in thermal equilibrium.  This leads to an effective single Boltzmann equation
\begin{align}
\label{eq:RecombBoltz}
\frac{1}{\lambda} \frac{d Y_\mathcal{Q}}{d z }  = - S_{\rm I, bsf}(z) \left( 2 Y_\mathcal{Q}(z)^2  - \frac{A g_\mathcal{Q}^2 z^{3/2}(Y_c -  Y_\mathcal{Q} (z))  e^{-z \Delta} }{g_I}\right)\,,
\end{align}
where $\Delta = E_B/\MDM$ and $A = 90/( (2 \pi)^{7/2} g_{\rm SM}^* ) $.
The value of $z_I$ is  defined by imposing that the leading order term in the $1/\lambda \ll 1$ expansion of the solution $Y_\mathcal{Q}(z) \approx Y_\mathcal{Q}^0(z) +Y_\mathcal{Q}^1(z)/\lambda$ is comparable to the second order term. 
The leading order term is simply defined by the condition that the derivative of $Y_\mathcal{Q}(z)$ vanishes 
\begin{align}
Y_\mathcal{Q}^0(z) =A\,z^{3/2}\, \frac{g_{\mathcal{Q} }}{4 g_I} e^{- z \Delta} \left(\sqrt{\left( g_{\mathcal{Q} }^2+8 \frac{ z^{-3/4}\,Y_c }{A} g_I e^{z \Delta }\right)}- g_{\mathcal{Q} }\right) \,.
\end{align} 
Inserting the assumptions in eq.~(\ref{eq:RecombBoltz}), solving for $Y_\mathcal{Q}^1(z)$ and evaluating $Y_\mathcal{Q}^0(z_I)= Y_\mathcal{Q}^1(z_I)/\lambda$ defines $z_I$. Such equation can be simplified assuming $z\gg1$ and reads 
\begin{align}
z_I =\frac{1}{\Delta} \ln{\left( \frac{32 A g_\mathcal{Q}^2 \lambda^2 S_{\rm I, bsf}(z_I)^2  Y_c  }{\Delta^2  g_I  z_I^{5/2}}\right)}\,.
\end{align}
For a  typical value  $\Delta\equiv  E_B/\MDM \approx 10^{-3}$ we find $z_I \approx 10^5$,
which justifies our initial assumptions, since $z_f \approx 25$ and the annihilation has no effect at $z>10^4$.  
Now the second effective eq.~(\ref{eq:RecombBoltz}) which describes the recombination effect can be integrated in the same manner as the first and leads, after the appropriate asymptotic matching, to
\begin{align}
 Y_\mathcal{Q} (\infty) = \left(  2 \lambda \int_{z_I}^\infty \frac{S_{\rm bsf-stable}(z)}{z^2}dz + \frac{1}{Y_\mathcal{Q}^0(z_I)} \right)^{-1} \,\qquad
 Y_I (\infty) = \frac{1}{2} \left( Y_\mathcal{Q}(z_I) - Y_\mathcal{Q}(\infty)\right)\,.
\end{align}
Using this method we find that, in the models considered, the relic abundance of stable dark di-quark states is at most at the percent level of the abundance of free dark quarks at confinement. In conclusion, perturbative production of stable bound states negligibly affects the final dark matter relic abundance.

\footnotesize

\bibliographystyle{abbrv}
\bibliography{mybib}

\end{document}